\documentclass[%
reprint,
 superscriptaddress,
 preprintnumbers,
 nofootinbib,
 amsmath,amssymb,
 aps]{revtex4-2}
 \usepackage[hyperindex,breaklinks]{hyperref}% add hypertext capabilities
\hypersetup{
        colorlinks=true,
        linkcolor=blue,
        filecolor=magenta,
        urlcolor=blue,
        citecolor=magenta
}

\usepackage{epsfig}
\usepackage{graphicx}
\usepackage[vcentermath,enableskew]{youngtab}

\usepackage[utf8]{inputenc}
\usepackage[english]{babel}
\usepackage{makeidx}
\usepackage{tikz}
\tikzset{every picture/.style={line width=0.75pt}} %set default line width to 0.75pt
\tikzset{line/.style={thick, decorate, draw=black,}}
\usepackage{amsfonts}
\usepackage{enumerate}
\usepackage{mathrsfs}
\usepackage{tensor}
\usepackage[autostyle]{csquotes}
\usepackage{subfig}

\def\be{\begin{equation}}
\def\ee{\end{equation}}
\def\bea{\begin{eqnarray}}
\def\eea{\end{eqnarray}}
\newcommand{\nn}{\nonumber}

\newcommand{\ft}[2]{{\textstyle\frac{#1}{#2}}}
\def\ii{{\rm i}}
\def\vi{p_\infty}

\def\apjl{\ref@jnl{ApJ}}

\usepackage{amsmath,amssymb}
\usepackage{latexsym}
\usepackage{graphicx}
\usepackage{slashed}
\usepackage{soul}

 \usepackage[vcentermath,enableskew]{youngtab}
\usepackage{epsfig}

\def\be{\begin{equation}}
\def\ee{\end{equation}}
\def\bea{\begin{eqnarray}}
\def\eea{\end{eqnarray}}

\begin{document}
\title{Gravitational wave forms for extreme mass ratio collisions from supersymmetric gauge theories}
\author{Francesco Fucito, Jose Francisco Morales}
\email{fucito@roma2.infn.it, morales@roma2.infn.it}
\affiliation{Sezione INFN  ``Roma Tor Vergata" and Dipartimento di Fisica, Universit\`a di Roma ``Tor Vergata", Via della Ricerca
	Scientifica 1, 00133, Roma, Italy}
\author{and Rodolfo Russo}
\email{r.russo@qmul.ac.uk}
\affiliation{School of Mathematical Sciences,
 Queen Mary University of London, Mile End Road, E1 4NS London, United Kingdom}

\begin{abstract}
	We  study the wave form emitted by a particle moving along an arbitrary (in general open) geodesic of the Schwarzschild geometry. The mathematical problem can be phrased in terms of quantities in ${\cal N}=2$ supersymmetric gauge theories that can be calculated by using localization and the AGT correspondence. In particular through this mapping, the post-Newtonian expansion of the wave form is expressed as a double instanton sum with rational coefficients that resums all tail contributions into Gamma functions and exponentials. The formulae we obtain are valid for generic values of the orbital quantum numbers $\ell$ and $m$. For $\ell=2,3$ we check explicitly that our results agree with the small mass ratio limit of the wave forms derived in the Multipole Post-Minkowskian and the amplitudes approaches. We show how the so-called tails and tails of tails contributions to the wave form arise in our approach. Finally, we derive a universal formula for the soft limit of the wave form that resums all logarithmic terms of the form $\omega^{n-1} (\log \omega)^n$.
\end{abstract}

\maketitle
\tableofcontents
\section{Introduction}
\label{sec:intro}

In the past few years the traditional approaches to the computation of the dynamics of gravitationally interacting  systems (see \cite{Blanchet:2013haa} for a comprehensive review and citations therein) have been complemented by a host of new techniques coming from such disparate fields as scattering amplitudes
\cite{Bjerrum-Bohr:2018xdl,Cheung:2018wkq,Bern:2019nnu,KoemansCollado:2019ggb,Bern:2019crd,Bjerrum-Bohr:2019kec,Bern:2020gjj,DiVecchia:2021ndb,DiVecchia:2021bdo,Cristofoli:2021vyo,Herrmann:2021tct,Brandhuber:2021eyq,Bjerrum-Bohr:2021vuf,Bjerrum-Bohr:2021din,Adamo:2022qci,Brandhuber:2023hhy,Herderschee:2023fxh,Elkhidir:2023dco,Georgoudis:2023lgf,DiVecchia:2023frv,Brandhuber:2023hhl,DeAngelis:2023lvf}, effective field theories \cite{Kalin:2020mvi,Kalin:2020fhe,Liu:2021zxr,Bjerrum-Bohr:2022ows}, worldline quantum field theories \cite{Mogull:2020sak,Jakobsen:2021smu,Jakobsen:2021lvp,Jakobsen:2023ndj,Jakobsen:2022psy}, two dimensional conformal field theories \cite{Novaes:2014lha,CarneirodaCunha:2015hzd,CarneirodaCunha:2015qln,Amado:2017kao,Lencses:2017dgf,BarraganAmado:2018zpa,Novaes:2018fry,CarneirodaCunha:2019tia,Amado:2020zsr,BarraganAmado:2021uyw,Cavalcante:2021scq,daCunha:2021jkm}, supersymmetric quantum gauge theories \cite{Aminov:2020yma,Bianchi:2021xpr,Bianchi:2021mft,Bianchi:2022wku, Bianchi:2023sfs,Aminov:2023jve}, AGT correspondence \cite{Bonelli:2021uvf,Bonelli:2022ten,Consoli:2022eey,Fucito:2023afe,Bautista:2023sdf,DiRusso:2024hmd,Arnaudo:2024rhv}.
It is very useful to compare these different approaches in the overlapping regions of applicability both to check the consistency of the results and to gain a better understanding of the physical problem. The Post-Newtonian (PN) and Multipole Post-Minkowskian (MPM) approach apply when velocities are small  and gravitational interactions are weak; black hole perturbation theory requires a small ratio between the masses of the two components of the binary system, but it makes no assumption on the strength of the gravitational interactions itself. Finally, scattering amplitudes allow for finite velocities but require weak gravitational interactions and open trajectories ending on asymptotic states.

In this paper we apply black hole perturbation theory to the study of the gravitational waves  produced by a light particle scattered in the Schwarzschild geometry. The deformation produced by the particle motion can be viewed as a perturbation of the Schwarzschild geometry and at linear order  it is  described  by a confluent Heun like equation, see \cite{RegWheel,Zerilli:1970se} for the original papers and \cite{Chandrasekhar:1985kt,Novikov:1989sz} for a comprehensive treatment of the subject. The PN expansion of the gravitational waves produced by a circular (and nearly circular) motion has been  computed in \cite{Poisson:1993vp,Shibata:1994jx,Tagoshi:1993dm,Tagoshi:1994sm,Tanaka:1996lfd,Mino:1997bx,Fujita:2010xj,Fujita:2011zk,Fujita:2012cm} using the so called MST algorithm.
Here we extend these results to arbitrary trajectories (bounded or not). We use the framework put forward in \cite{Fucito:2023afe}, where the confluent Heun equation is realized as the quantum version of a Seiberg-Witten elliptic curve describing the low energy dynamics of a gauge theory with ${\cal N}=2$ supersymmetry. The solution of the Teukolsky (or Regge Wheeler) equation is then computed  by the instanton partition function of a quiver gauge theory with gauge group $SU(2)^2$ and couplings $q_1=2  M/r$ and $q_2=2i\omega r$. Interestingly, the PN limit, where distances are large and velocities are small, corresponds precisely to the weak coupling limit $q_1,q_2\ll 1$  of the gauge theory and can be computed using localization techniques.  This produces remarkable compact formulae  for the wave form components of generic values of the orbital harmonic degree $\ell$,  where the infinite towers of tail and tail-of-tail contributions are resummed into Gamma functions and exponentials.

 We first compute the PN wave form, in the time and frequency domains, produced by a particle moving along an arbitrary (in general open) trajectory in a Schwarzschild geometry. In this way we reproduce the results obtained in literature using the MPM expansion and generalize them to arbitrary values of $\ell$.
Already at this stage the formalism provides a partial resummation of all the contributions dubbed as gravitational wave tails and tails of tails. We then apply the results to the study of hyperbolic collisions and compare it against  \cite{Bini:2023fiz,Georgoudis:2023eke,Georgoudis:2024pdz,Bini:2024rsy} obtained via the MPM and  scattering amplitude techniques.

Finally we study the soft limit of the gravitational wave emission during a heavy light scattering. A renewed interest in the soft graviton theorem \cite{Weinberg:1965nx} has been recently fostered by a series of papers \cite{Strominger:2013jfa,Strominger:2014pwa,He:2014laa,Cachazo:2014fwa} in which the leading soft and subleading soft gravitational theorems were linked to the Ward identities associated to the algebra of an infinite dimensional asymptotic symmetry: BMS supertranslations and superrotations.

The leading contribution $c_0 \omega^{-1}$ of this amplitude is known from the famous soft graviton theorem \cite{Weinberg:1965nx}. The coefficient $c_0$ is universal and depends only on the momenta of the incoming and outgoing particles. In \cite{Chakrabarti:2017ltl,Chakrabarti:2017zmh,Sahoo:2018lxl,Saha:2019tub,Sahoo:2020ryf,Sahoo:2021ctw,Hait:2022ukn} subleading corrections of the form $c_n   \omega^{n-1}(\log\omega)^{n}$ were studied and explicit results for $c_1$ and $c_2$ were obtained for general kinematics. In~\cite{Alessio:2024onn} it was conjectured that the whole series of soft terms labelled by $c_n$ above is also universal, in the sense that it is determined by the initial and the final state of the hard particles. Here we prove this conjecture for the case of $2 \to 2$ scattering in the limit in which the ratio of the two masses is small (probe limit) and derive a formula for the soft wave form which is exact in the other parameters. The result resums the infinite tower of logarithmic terms $c_n   \omega^{n-1}(\log\omega)^{n}$  into an exponential finding perfect agreement with the conjecture of~\cite{Alessio:2024onn}.

The plan of the paper is the following: in Section~\ref{sec:bhp} we discuss some introductory material. In Section~\ref{sec:wf} we discuss the waveform exploiting the results in \cite{Fucito:2023afe} and rewrite them in the frequency and time domain. In Section~\ref{sec:hyp} we discuss the general case of hyperbolic trajectories. Finally in Section~\ref{sec:soft} we discuss the soft limit. In Section~\ref{summary} we summarize our results and highlight some possible developments that we find interesting. Some technical material is confined to the appendices. While we were writing this paper, other studies \cite{Alessio:2024onn,Bini:2024ijq,Sen:2024bax} dealing with the soft limit appeared on the arXiv and have a partial overlap with our results in Section~\ref{sec:soft}.

\section{Black hole perturbations}
\label{sec:bhp}

In this section we review the basic ingredients entering the description of the linear perturbations of a Schwarzschild geometry generated by the motion of a massive particle.

\subsection{Geodetic motion}

We consider  the motion of a particle of mass $\mu$ in the Schwarzschild  metric
\be
ds^2= -f(r) dt^2 + {  dr^2\over f(r) } + r^2 (d\theta^2+\sin^2\theta d\phi^2 )\;,
\ee
with
\be
 f(r) =1-{2M\over r}\;.
\ee
 $M$ is the product of the Newton constant $G$ and the black hole mass, $M_{\rm bh}$ i.e. a length rather than a mass scale\footnote{ We use the mostly plus signature and keep $G$ different from one, see~(\ref{teukbis}).}. In order to restore the mass meaning of $M$, it is sufficient to send $M \to G \,  M_{\rm bh}$ in all our formulae.

We work in the Hamiltonian formalism where the geodesic motion is governed by the Hamiltonian
 \be
 {\mathcal H}=\ft12  g^{\mu\nu} p_\mu p_\nu = -{\mu^2 \over 2}\;.
 \ee
We introduce the affine parameter $d\tau=\mu^{-1}\sqrt{g_{\mu\nu} dx^\mu dx^\nu}$, normalised such that  the quadri-velocity $u^\mu$ and momentum $p^\mu$ coincide
\be
u^\mu= p^\mu=\frac{d x^\mu}{d\tau} ={\partial {\cal H} \over \partial p_\mu}\;.
\ee
We denote by $E=-p_t$ and $J=p_\phi$ the conserved energy and angular momentum.
In terms of these variables the geodesic equations become
\begin{equation}\label{uvel}
\begin{gathered}
\frac{dt}{d\tau}={E\over f(r) }  \quad, \quad \frac{d\phi}{d\tau}={J\over r^2 } \, \quad, \quad \frac{d\theta}{d\tau}=0 \\
 \frac{dr}{d\tau}  =  \sqrt{  E^2    - f(r) \left(  \mu^2+ {J^2 \over r^2}   \right)  }   \;,
\end{gathered}
\end{equation}
or equivalently
\begin{equation}\label{eq:dert}
  \begin{gathered}
     \dot \phi= {f(r)\over E} {J\over r^2 }  \qquad , \qquad \dot{\theta}=0\;,\\  \dot{r}= {f(r)\over E}  \sqrt{  E^2    - f(r) \left(  \mu^2+ {J^2 \over r^2}   \right)  }\;,
  \end{gathered}
\end{equation}
where the dot stands for a time derivatives.
Without loss of generality we can set $\theta=\pi/2$ as the orbital plane. Alternatively one  can specify the trajectory by its asymptotic momentum $p$
and the impact parameter $b$  related to
 the energy  and momentum  via
\be
E=\sqrt{\mu^2+p^2} \qquad, \qquad J=p\,  b\;.
\ee
The stress energy tensor generated by a particle moving along the trajectory $x(\tau)$ reads
\bea
{\cal T}^{\mu\nu} &=&  \int d\tau  {u^\mu u^\nu  \over r^2 \sin\theta}  \delta^{(4)} \left[ x- x(\tau) \right]\;.
 \eea
   We are interested in the  gravitational wave generated  as a back reaction  of the stress energy source on the geometry.  We will denote by $\omega$ the frequency of the emitted wave.
We will mainly consider three special limits according to the relative scales defined by $M$, $\omega$ and $J/E$.

\begin{itemize}

\item{ Post-Newtonian (PN) expansion: In this limit we consider scattering at small velocities  and distances large compared to the Schwarzschild radius  but still small with respect to the wavelength of the emitted wave.
To trace the PN order we introduce the bookkeeping parameter $\eta$ and scale velocities with $\eta$ and $M$ with $\eta^2$ in such a way that the kinetic and potential energies scale in the same way, i.e.
   \bea
{\rm PN}:\quad &&\, p \to p\, \eta     ,\, ~  M \to M \, \eta^2  ,\, ~
\omega  \to \omega \, \eta   ,\nn\\&&   J\to J \, \eta   , \, ~ \eta \to 0\;. \label{ejmwPN}
\eea
The powers of $\eta$ conventionally account for corrections of  half PN order. }
\item{ Post-Minkowskian (PM) expansion:  in this limit $G\to0$ and therefore
  \be
  {\rm PM}:\qquad  M \to 0\;.
  \ee
  }
  \item{ Soft expansion:  In this limit, the wave length  is larger than all other distance scales involved
  \be
  {\rm Soft}:\qquad   \omega \to 0\;.
  \ee
  }
  \end{itemize}

   In the case of bounded quasi-circular orbits, the PN and PM limits are related since gravitational potentials and kinetic energies are in the average of the same order. In addition, the frequency $\omega$ is determined
 in terms of the orbital angular velocities so also the wave length scale matches the one settled by the gravitational and kinetic energies.
 For more general orbits instead, the kinematics is described by three independent parameters that can be tuned at will.

\subsection{Teukolsky equation  }

We consider the perturbations of the metric generated by the motion of a light particle in the Schwarzschild geometry.
Following \cite{Teukolsky:1972my}, the metric perturbations can be conveniently decomposed according to their helicity $s$, with the physical components $h_+$, $h_\times$
  related to the projection of the Weyl tensor with helicity $s=-2$, the so called $\psi_4$ mode. We refer the reader to Appendix \ref{apTeu} for notations and details.

    At linear order, the Einstein equation for $\psi$ can be separated into a radial and an angular part via the ansatz
\begin{equation}
\psi (X) = R^4 \psi_4(X)= \int\!  {d\omega\over 2 \pi} \sum_{\ell,m} e^{-{\rm i}\omega T}  R_{\ell m}(R)   Y_{-2}^{\ell m} (\Theta,\Phi)\;,
 \label{psibis}
\end{equation}
with $X=(T,R,\Theta,\Phi)$ the coordinates of the observer and $ Y_{-2}^{\ell m} (\Theta,\Phi)$  the spin weighted spherical harmonic functions.
We use capital letters for the coordinates of the observer to distinguish them from the coordinates $x^\mu=(t,r,\theta,\phi)$ that will be used to describe the trajectory of the light particle.
The radial function $R_{\ell m}(r)$  satisfies the Teukolsky equation \cite{Teukolsky:1972my}
\begin{widetext}
\begin{equation} r^4 f(r)^2 {d\over dr} \left[   { R_{\ell m}'(r) \over r^2 f(r) } \right] { +} \left(\frac{\omega^2 r^2{+}4{\rm i} \omega (r-M) }{f(r)}{-}8{\rm i}  \omega r{-}(\ell{+}2)(\ell{-}1) \right)R_{\ell m}(r)  = G  T_{\ell m}(r)\;, \label{teukbis}
\end{equation}
\end{widetext}
 with $T_{\ell m}(r) $ the harmonic decomposition of the stress energy source.
 The solution of the homogeneous equation (\ref{teukbis}) can be obtained with the Green function method  and written in the form
 \be \label{eq:R-Z}
 R_{\ell m }(R)  = \mathfrak{R}_{ {\rm up},\ell } (R) \,Z_{\ell m}(\omega)\;,
\ee
 where
 \be
  Z_{\ell m}(\omega)=     \int_{2M }^\infty  \mathfrak{R}_{{\rm in},\ell} (r')     {T_{\ell m }(r') \over  (r')^4 f(r')^{2}  } dr' \label{rt0bis}
 \ee
 and  $\mathfrak{R}_{ {\rm in},\ell }(r)$, $\mathfrak{R}_{ {\rm up},\ell }(r)$ are solutions of the homogeneous equation, such that $\mathfrak{R}_{ {\rm in},\ell }(r)$ satisfies incoming boundary conditions at the horizon and $\mathfrak{R}_{ {\rm up},\ell }(r)$
 outgoing ones at infinity.  We normalize $\mathfrak{R}_{ {\rm in},\ell }(r)$, $\mathfrak{R}_{ {\rm up},\ell }(r)$ in such a way that the Wronskian is given by
 \be
W= \mathfrak{R}_{ {\rm in},\ell }(r) \mathfrak{R}_{ {\rm up},\ell }'(r) -\mathfrak{R}'_{ {\rm in},\ell }(r) \mathfrak{R}_{ {\rm up},\ell }(r)= r^2  f(r)
 \ee
and consequently, the Green function is given by  the radially ordered product of the two solutions.  This can be achieved by requiring the asymptotic behaviour\footnote{Using the Teukolsky equation for $s=-2$, it is easy to prove that the ratio $W/(r^2 f)$ is constant. The constant can be evaluated at $r\to \infty$ using (\ref{rinf}).}
\begin{align} \label{rinf}
  \mathfrak{R}_{ {\rm in},\ell }(r) & \underset{r\to \infty}{\approx} B r^3 e^{{\rm i} \omega r_*}  + { e^{-{\rm i} \omega r_*} \over 2{\rm i} \omega r }  \,,
 \\ \mathfrak{R}_{ {\rm up},\ell }(r) & \underset{r\to \infty}{\approx} r^3 e^{{\rm i} \omega r_*} \label{rinf2}
\end{align}
 with $r_*=r+2M \log \ft{r}{2M}$ and $B$ some constant.  We notice that  due to the spherical symmetry    $ \mathfrak{R}_{{\rm in},\ell} (r)$, $\mathfrak{R}_{{\rm up},\ell} (r)$ depend only on the orbital number $\ell$ and not on $m$.

    The precise form of the source term $T_{\ell m}(r)$ and the  harmonic coefficients $Z_{\ell m}$ for the motion in bounded orbits of Schwarzschild and Kerr geometries were derived in  \cite{Poisson:1993vp,Tagoshi:1993dm,Shibata:1994jx,Tagoshi:1994sm,Tanaka:1996lfd,Mino:1997bx}. In Appendix \ref{apTeu}, we review the computation of $T_{\ell m}(r)$ and $Z_{\ell m}(\omega)$ and generalize it to open trajectories in Schwarzschild geometry. In Appendix \ref{appendixC}, we review the  quiver gauge theory representation of the solutions of  the confluent Heun equation.
    The results read
    \be
   Z_{\ell m }(\omega)  =  \int\!  dt \, e^{-im\phi(t)+i\omega t} {  1 \over   r(t) ^2 } \sum_{i=0}^2  b^i_{\ell m}  {\cal L}^i \, \left[  \mathfrak{R}_{{\rm in},\ell}(r(t) ) \right],  \label{za0}
\ee
 where
\bea
{\cal L}^0 &=&   { E\over f}  \left(1+{\dot{r} \over f}  \right)^2\;,  \nn    \\
 {\cal L}^1 &=&      {  {\rm i}  J  \over r}  \left(1+{\dot{r} \over f}  \right)    \left(2 -r \partial_r  +{ {\rm i}  \omega r \over   f(r)  }\right)\;,  \label{ls}  \\
 {\cal L}^2 &=& {J^2   \over  E    } { f(r) \over  r^2}  \left[     f(r)^{-2}  \left( {-} {\rm i}   \omega r  {+}  \ft{1}{2}  \omega^2 r^2{+} {\rm i} \omega M     \right)\right.\nn\\&& \left.{+}  \left(1{+}{{\rm i}  \omega r \over f(r) }\right) r \partial_r    {-}\ft12    r^2 \partial^{2}_r \right]\;,     \nn
\eea
 and
  \bea
 b^0_{\ell m}  &=& {  \pi   \over 2   }   \sqrt{(\ell-1)\ell(\ell+1)(\ell+2)} \,   Y_0^{\ell m}(\ft{\pi}{2},0)\;,    \nn\\
b^1_{\ell m}  &=&     \pi     \sqrt{(\ell-1)(\ell+2) }   Y_{-1}^{\ell m} (\ft{\pi}{2},0)\;,    \label{bbb} \\ \nn
 b^2_{\ell m}  &=&   \pi   \, Y_{-2}^{\ell m} (\ft{\pi}{2},0)\;.
\eea
  For later use, we record the identities satisfied by the coefficients above
  \begin{equation}
    b^1_{\ell m} = {2m \over \ell(\ell{+}1)   } b^0_{\ell m}   ,    b^2_{\ell m} = -{2(\ell^2{+}\ell{-}2 m^2) \over \ell(\ell^2{-}1)(\ell{+}2)   } b^0_{\ell m}\;,
  \end{equation}
  valid for $\ell+m\in 2\,\mathbb{Z}$ and
  \begin{equation}
   b_{\ell m}^2  =   { 2m\,  b^1_{\ell m}  \over  (\ell{-}1)(\ell{+}2)   }   , ~~ b^0_{\ell m} =0\;,
  \end{equation}
valid for $\ell{+}m\in 2\,\mathbb{Z} {+}1$.

\section{The wave form}
\label{sec:wf}

 In this section we compute the PN expansion of the gravitational wave form produced by the motion of a particle along an arbitrary trajectory in the Schwarzschild geometry.
We consider linear perturbations $g_{\mu\nu}=g_{\mu\nu}^{\rm Sch}+h_{\mu\nu}$ produced by the particle motion, and write
 \be
 \label{eq:hX0}
h(X) = h_{+}(X) -{\rm i}  h_{\times} (X)=h_{ij} (X)  {\bf e}_i {\bf e}_j\;,
 \ee
 where the polarization vector is
  \begin{align}
 {\bf e}&=\ft{1}{\sqrt{2}}\left( \partial_\Theta {\bf n} -{\rm i} {\partial_\Phi {\bf n}\over \sin\Theta}  \right)\\ &=\! \ft{1}{\sqrt{2}} \!\left(\cos\Theta \cos\Phi+{\rm i} \sin\Phi, \cos\Theta \sin\Phi-{\rm i} \cos\Phi,-\sin\Theta\right)
 \label{polarizationvector}
 \end{align}
  and the unit vector pointing from the origin to the observer is
\be
 {\bf n} = (\sin \Theta \cos\Phi , \sin \Theta \sin\Phi, \cos\Theta)\;.  \label{unitvector}
 \ee
 The wave form $h(X)$ is related to the spin two component $\psi_4$  by the standard  relation\footnote{The extra sign with respect to
\cite{Teukolsky:1972my} arises from the fact that we work in the mostly plus signature, see also (\ref{psi4}).}
 \be
\psi_4 (X) =- \ft12 \ddot{h}(X)    \label{psi4hdot}
\ee
 that relates the wave form $h(X)$  in the time domain to the fundamental mode $\psi_4(X) $.
 Expanding in harmonics, and combining (\ref{psibis}), (\ref{eq:R-Z}) and (\ref{psi4hdot}) one finds
\begin{equation}\label{eq:hX}
 h(X)   \underset{R\to \infty}{\approx} {4 G\over R} \int  {d\omega\over 2 \pi} \, \sum_{\ell,m} e^{-{\rm i}\omega (T-R_*)}       \,W_{\ell m}(\omega)     Y_{-2}^{\ell m} (\Theta,\Phi)\;,
\end{equation}
 with
 \be
   W_{\ell m}(\omega)=- {Z_{\ell m }(\omega) \over 2({\rm i} \omega)^2} =   \int  dt \, e^{-im\phi(t)+i\omega t} {\cal A}_{\ell m}(\omega, r(t))     \label{za}
\ee
 and
 \be
{\cal A}_{\ell m} (\omega, r)=- {  1 \over 2 ({\rm i} \omega r )^2 } \sum_{i=0}^2  b^i_{\ell m}  {\cal L}^i \, \left[  \mathfrak{R}_{{\rm in},\ell}(r) \right] ~. \label{alm2}
 \ee
 In the following we will split the contributions  ${\cal A}_{\ell m}$ according to whether
   $\ell+m$ is even or odd. They will be later related to the so called $U$  and $V$ radiative multipoles, so we will consequently name the corresponding contributions as
   \be
{\cal A}_{\ell m} =
\left\{
\begin{array}{lll}
{\cal A}^U_{\ell m}    & ~~~  &  \ell+m\in 2\,\mathbb{Z}  \\
 {\cal A}^V_{\ell m} &   &  \ell+m\in 2\,\mathbb{Z} +1 \\
\end{array}
\right.
\ee

\subsection{The  homogeneous solution}

The basic ingredient in the computation of the wave form is  the solution $\mathfrak{R}_{{\rm in},\ell}$ of the homogeneous equation with incoming boundary conditions. The PN expansion of this solution has been studied in the past \cite{Poisson:1993vp,Tagoshi:1993dm,Shibata:1994jx,Tagoshi:1994sm,Tanaka:1996lfd,Mino:1997bx} computing $\mathfrak{R}_{{\rm in},\ell}$ recursively order by order in $\ell$.
More recently \cite{Bonelli:2021uvf,Bonelli:2022ten,Consoli:2022eey,Fucito:2023afe,Bautista:2023sdf,DiRusso:2024hmd,Arnaudo:2024rhv}, these results were reproduced by a gauge theory inspired localization technique whose connection with gravity is provided by the AGT correspondence \cite{Alday:2009aq} which states that the partition function of the ${\cal N}=2$ supersymmetric gauge theory is related to the conformal blocks of a two dimensional conformal field theory. In turn the latter are connected to the solutions of a differential equation \cite{Belavin:1984vu} which can be mapped to the Teukolsky differential equation. Therefore the partition function of the ${\cal N}=2$ supersymmetric gauge theory provides a combinatorial formula
for the PN expansion of $\mathfrak{R}_{{\rm in},\ell}$  for generic values of $\ell$. The results are written as (see Appendix \ref{appendixC} for details)
 \be
 \mathfrak{R}_{\rm in,\ell}  =  \alpha_\ell  \,  ( {\rm i} \omega r)^{\ell+2}\,  \mathfrak{R}^{\rm tail}_{\ell} (\omega ) \,  \mathfrak{R}^{\rm flow}_{\ell} ( \omega, r) \,  \sum_{n=0}^\infty \mathfrak{R}^{\rm inst}_{ n}(\omega ,r)\;,
 \label{pnexpbis}
 \ee
with
  \begin{align}
  \label{p2final}
\alpha_\ell & =  (-2)^{ \ell+3  }
{ \Gamma(\ell-1 ) \over   \Gamma(2\ell+2 ) }\;,  \\
    \mathfrak{R}^{\rm flow}_{\ell} ( \omega, r)  & =  (-2{\rm i} \omega r )^{  \widehat{\ell}-\ell}   \nn \\
\mathfrak{R}^{\rm tail}_{\ell} (\omega  )  &= ({-}4{\rm i} \omega M)^{ 2{\rm i} \omega M}{ \Gamma(2\ell{+}2 )   \over   \Gamma(2\widehat{\ell}{+}2 )  } {    \Gamma(\widehat{\ell}{-}1{-}2{\rm i} \omega M ) \over     \Gamma(\ell{-}1 ) } \nn
\end{align}
where $\hat\ell(\omega)$ is connected to the Seiberg-Witten period \cite{Consoli:2022eey} and to the renormalized angular momentum of \cite{Sasaki:1981sx} as verified in a different context also in \cite{Bianchi:2024vmi}. For small $\omega r$ and $\omega M$ the various ingredients in (\ref{p2final}) can be expanded as
\begin{align}
    \mathfrak{R}^{\rm flow}_{\ell} ( \omega, r)  & =   1 + (\widehat{\ell} - \ell) \log(-2{\rm i} \omega r ) +\ldots,  \nn\\
\mathfrak{R}^{\rm tail}_{\ell} (\omega  )  &= 1{+}2 {\rm i} \omega M \left[  \log({-}4 {\rm i} \omega M){ -}\psi(\ell-1)  \right] {+}\ldots\;,  \nn\\
\hat{\ell} (\omega) &=\ell+ \beta^{[2]}_\ell \,\omega^2 M^2+ \beta^{[4]}_\ell \,\omega^4 M^4 + \ldots \label{lhatl}
\end{align}
with
\begin{align}
&\beta^{[2]}_{\ell}   = -2 \frac{\left(15 \ell^4+30 \ell^3+28 \ell^2+13 \ell+24\right) }{\ell (\ell+1) (\ell-1) (2 \ell+1) (2 \ell+3)}\;,  \label{beta2def} \\
&\beta_\ell^{[4]}  = {-}2 \Big[
 18480 \ell^{16}{+}147840 \ell^{15}{+}456120 \ell^{14}{+}605640 \ell^{13}\nn\\&{+}8295 \ell^{12}{-}1096830
   \ell^{11} {-}1678310 \ell^{10} {-}1520455 \ell^9{-}1355518 \ell^8\nn\\
   &{-}1397512 \ell^7{-}1217380 \ell^6{-}733273 \ell^5{+}675625 \ell^4 { +}1855326 \ell^3\nn\\&{+}850608 \ell^2
   {-}102816 \ell{-}51840 \Big] \Big[(\ell-1) \ell^3 (\ell+1)^3\nn\\& (\ell+2) (2 \ell-3) (4 \ell^2-1)^3   (2\ell+3)^3 (2 \ell+5)\Big]^{-1} \label{beta4def}
\end{align}
  The first term in the expansion~\eqref{lhatl} is relevant for the tail of tail contribution to the wave form as discussed in Sect.~\ref{sec:tails}.
Finally  $\mathfrak{R}^{\rm inst}_{n}$ collects  the instanton corrections to the quiver gauge theory partition function  scaling as $\eta^n$ (PN order $\ft{n}{2}$-th). The first few coefficients are
\begin{align}
\mathfrak{R}^{\rm inst}_{ 0} &= 1\;, \nn\\
\mathfrak{R}^{\rm inst}_{ 1} &= \frac{2 i  \omega r  }{1+\ell}\;, \nn\\
\mathfrak{R}^{\rm inst}_{ 2} &= -\left(\ft{\ell}{2}+1\right)  \ft{2M}{r} -\frac{(9+\ell) (\omega r) ^2}{2 (\ell+1) (2 \ell+3)}\;, \label{gothicR} \\
\mathfrak{R}^{\rm inst}_{ 3} &= -\frac{i (7+3 \ell) (2\omega M) }{2 (1+\ell)}-\frac{i (4+\ell) (\omega r) ^3}{(\ell+1) (\ell+2) (2 \ell+3)}\;, \nn\\
\mathfrak{R}^{\rm inst}_{4} &= \frac{-4 M^2(\ell^2-1) (\ell+2) }{(4-8 \ell) r^2}+\frac{(\ell+4) (\ell+9)(2\omega^2 r M) }{4 (\ell+1) (2 \ell+3)} \nn \\ & + \frac{\left(50+19 \ell+\ell^2\right) (\omega r)^4}{8 (1+\ell) (2+\ell) (3+2 \ell) (5+2 \ell)}\;. \nn
\end{align}
More terms can be found in \cite{Fucito:2023afe}. In the two limits $\omega r \ll M/r$ and $M/r \ll \omega r $ the double series can be resummed into hypergeometric functions
\begin{align}
\sum_{n=0}^\infty \mathfrak{R}^{\rm inst}_{ n}\!\!\! &\underset{\omega r \ll M/r } {=}\!\!\! \left( 1-\ft{2M}{r} \right)^{-{1\over 2}}  {}_2F_1\left(-\ell-2,-\ell;-2 \ell;\frac{2 M}{r}\right),  \nn\\ \label{eq:larger}
  \sum_{n=0}^\infty \mathfrak{R}^{\rm inst}_{ n}\!\!\!  &\underset{M/r \ll \omega r } {=} \!\!\! e^{- {\rm i} \omega r  } \, _1F_1(\ell+3;2 \ell+2;  2{\rm i} \omega r )\;.
  \end{align}
In these limits $\mathfrak{R}^{\rm tail}_{\ell} (\omega )   \approx \mathfrak{R}^{\rm flow}_{\ell} ( \omega,r ) \approx 1$. We will exploit this result in Sect.~\ref{ssec:leadlogsof} when discussing the soft limit of the waveform.

\subsection[The PN expansion of A (omega ,r)]{The wave form in the frequency domain}

In this section we display the first terms in the PN expansion of ${\cal A}_{\ell m}$ defined in~\eqref{alm2} that determines the waveform coefficients $W_{\ell m}$ upon integration (\ref{za}). We write explicit formulae up to 1.5PN beyond  the leading order, but the generalization to higher orders is straightforward.
We start by expanding (\ref{alm2})  and use partial integration to rewrite all terms involving time derivatives $\dot{r}$ in terms of $r$ and $J$.
   After dropping total derivative terms\footnote{For instance the ${\cal O}(\eta)$ correction (beyond the leading order)  to~\eqref{aulm} coming from the first subleading terms in ${\mathcal L}^0$ and $\mathfrak{R}^{\rm inst}_1$ combine with the leading contribution from the term ${\mathcal L}^1$ to produce a total derivative that cancels upon integration. Similarly, the first subleading terms in ${\mathcal L}^1$ and $\mathfrak{R}^{\rm inst}_1$ combine with the leading contribution from the term ${\mathcal L}^2$ to produce a total derivative  term in~\eqref{avlm} that cancels upon integration.} that cancel upon integration in~\eqref{za}, one finds up to 1.5PN beyond the leading order
\bea \label{aulm}
&& {\cal A}^{\rm U}_{\ell m}   = \eta^\ell \, E\, c_{\ell m}^0\,    ( {\rm i} \omega r )^{\ell}\,  \Big\{  1{+}
\eta^2 \Big[\frac{J^2 ( \ell^2{-}m^2)}{\ell (\ell{+}1) E^2
   r^2}\\
   &&{-}\frac{2 J (\ell{-}1) m \omega }{\ell (\ell{+}1) (\ell{+}2) E}{-}\frac{\left(\ell^2{+}\ell{-}1\right) M}{(\ell{+}1) r}  {-}\frac{(\ell{-}1) \ell r^2
   	\omega ^2}{2 (\ell{+}1) (\ell{+}2) (2 \ell{+}3)}\Big] \nn\\
&&   {-} 2{\rm i} M \omega  \eta^3 \Big[\kappa_\ell - \gamma+\frac{{\rm i}\pi}{2} {-} \log (4 M  \omega)
    \Big]+\ldots  \Big\}\;, \nn\\
&& {\cal A}^{\rm V}_{\ell m}   =  \eta^{\ell+1}  J\, \ell \, c_{\ell m}^1\, \omega    ( {\rm i} \omega r  )^{\ell-1}\,  \Big\{1 {-}\eta^2 \Big[\frac{J m \omega }{E (\ell{+}1) (\ell{+}2)}\nn\\&&{+}\frac{(\ell{-}1) (\ell{+}2)
   M}{\ell r}{+}\frac{(\ell{-}1) r^2 \omega ^2}{2 (\ell{+}1) (2 \ell{+}3)}\Big] \label{avlm} \\  \nn
   &&  ~~~~~~  -2{\rm i} M
   \eta^3 \omega  \Big[  \pi_\ell  - \gamma+\frac{{\rm i}\pi}{2} {-} \log (4 M \omega ) \Big]+\ldots  \Big\}\;,
 \eea
 with
 \bea
 c^i_{\ell m} &=& -{\alpha_\ell \, b^i_{\ell m}\over 2} ,\qquad  \kappa_\ell=\frac{ 2\ell^2{+}5\ell{+}4}{\ell(\ell{+}1)(\ell{+}2)} {+}{1\over 2} {+}
 H_{\ell-2},\nn\\ \pi_\ell&=& \frac{  \ell-1}{\ell(\ell+1)} {+}{1\over 2} {+} H_{\ell-1}\;, \eea
 where $H_n =\sum_{k=1}^{n} \ft{1}{k}$ are the harmonic numbers and $\gamma$ is Euler's constant. The ${\cal O}(\eta^3)$ terms above are the leading tail corrections~\cite{Blanchet:1992br} written in the frequency domain, see for instance Eq.~(79) of~\cite{Georgoudis:2024pdz}. In the result above there is a non-observable additive shift\footnote{This shift was already pointed out in~\cite{Poisson:1993vp,Poisson:1994yf} and is due to the use of Schwarzschild, instead of harmonic, coordinates, as discussed in~\cite{Blanchet:1993ec}. We will have to pay attention to the coordinate system used also in Sect.~\ref{sect:multmom} when comparing the source multipole moments with the results in the literature.} by a factor of $1/2$ with respect to usual formulae of~\cite{Blanchet:2013haa}. This can be absorbed (together with other real $\ell$-independent shifts) in a redefinition of the origin of the retarded time. We wrote explicitly the factors of $\log(-{\rm i})=-\frac{{\rm i}\pi}{2}$: even if they are $\ell$-independent, it is not possible eliminate them with a redefinition of the retarded time since they are imaginary. These tail contributions were derived in the context of black hole perturbation theory in~\cite{Poisson:1994yf}, but with a different technical approach.  For later use let us write explicitly the $\ell=2$ modes\footnote{The tail contribution for the $\ell=m=2$ mode was first derived in~\cite{Anderson:1982fk} which also obtained the tail of tail coefficient $\beta^{[2]}_{2}$, see~\eqref{beta2def}.}
  \begin{align}
  {\cal A}^{\rm U}_{2,\pm2} &=  {\eta^2\over 2} E\sqrt{\ft{\pi}{5}} ({\rm i}  \omega r)^2 \left[    1-\eta ^2 \left(\frac{5 M}{3 r}\pm \frac{J \omega }{6 \mu }+\frac{r^2 \omega^2}{84}\right)\right. \nn \\ & \left. -2 i M \eta ^3 \omega  \left(\frac{17}{12}-\!\gamma +\frac{i \pi }{2}\!-\log \left(4 M \eta ^3 \omega \right)\right) \right], \\
  {\cal A}^{\rm U}_{20} & =  {-}  \eta^2 E\sqrt{\ft{\pi}{30}} ( {\rm i} \omega r )^2 \left[   1{-}\eta ^2 \left(\frac{5 M}{3 r}{-}\frac{2 J^2}{3 r^2 \mu ^2}{+}\frac{r^2 \omega^2}{84}\right) \right. \nn \\ & \left. - 2 i M \eta ^3 \omega  \left(\frac{17}{12}{-}\gamma {+}\frac{i \pi }{2}{-}\log \left(4 M \eta ^3 \omega \right)\right) \right]\;,  \\
     {\cal A}^{\rm V}_{2,\pm1} &= { \pm   \frac{2{\rm i}\eta^3 }{3} }  \sqrt{\ft{\pi}{5}} J \omega^2 r   \left[  1 {-}\eta ^2 \left(\frac{2 M}{r}\pm \frac{J \omega }{12 e}{+}\frac{r^2 \omega^2}{42}\right) \right. \nn \\ & \left. - 2 {\rm i} \omega M \eta ^3 \left(\frac{10  }{6}{-}\gamma {+}\frac{i
   \pi }{2}{-}\log \left(4 M \eta ^3 \omega \right) \right) \right]\;. \label{a2m}
\end{align}

\subsection{The wave form in the time domain}

The wave form in the time domain is defined by the Fourier transform
\be
 W_{\ell m} (u)=     \int  {d\omega\over 2\pi}  dt \, e^{-im\phi(t)+i\omega (t-u) } {\cal A}_{\ell m}(\omega, r(t))\;,   \label{fourier}
\ee
where ${\cal A}_{\ell m}$ is given by (\ref{alm2}).
 It is convenient to split ${\cal A}_{\ell m}$ into instantaneous and  tail contributions
\be
{\cal A}_{\ell m}(\omega,r)={\cal A}^{\rm inst}_{\ell m}(\omega, r){\cal A}^{\rm tail}_{\ell m}(\omega, r)\;,
\ee
with ${\cal A}^{\rm tail}_{\ell m}(\omega, r)=( 1+{\cal B}^{\rm tail}_{\ell}(\omega,r))$.
The first factor in the r.h.s. collects contributions depending polynomially on $\omega$ which are the so called instantaneous contributions of the multipole moments to the wave form.
The tail part contains $\log\omega$ terms and produces an infinite tower of tails and tail-of-tail terms in the language of \cite{Blanchet:2013haa}.
  In this section we focus on instantaneous terms, leaving for the next section the study of tail and tail of tails contributions.

  The instantaneous part will  be written
 in the simple form
\be
{\cal A}^{\rm inst}_{\ell m} (\omega,r(t) )=({\rm i} \omega)^\ell \widehat{\cal A}^{\rm inst}_{\ell m} (t) \label{hatalm}
\ee
 after getting rid of any extra powers of $\omega$  in (\ref{fourier}) by integrating them by parts.
      Integration by parts translate power of $\omega$'s  into time derivatives of the integrand that
 can be computed using the geodesic equations.
 More precisely, the energy and angular momentum can be written in terms of radial and angular velocities
\bea
E &=&\mu\left( 1+\eta^2\left[ -{M\over r}+{\dot{r}^2\over 2}+{ r^2 \dot{\phi}^2\over 2}  \right]+\ldots  \right),\nn\\  J&=&{E  r^2 \over f(r)  } \dot{\phi}\;, \label{ejdot}
\eea
 while higher time derivatives can be computed using
   \be
  \ddot{\phi}=-{2 \dot{r} \dot{\phi} \over r} +\ldots
 \qquad , \qquad \ddot{r}=- {M \over r^2} +r \dot{\phi}^2   +\ldots
\ee
that follows from (\ref{ejdot}) after taking a time derivative.  Proceeding in this way, one finds  up to the 1PN order
 \begin{align}\label{almtime}
   \widehat{\cal A}^U_{\ell m}(t)  = \mu & c_{\ell m}^0   r ^\ell \left\{ 1+\eta^2 \left[-\frac{ {\rm i} (\ell-1) (\ell+3) m r \dot{r}\dot{\phi} }{(\ell+1) (2 \ell+3)} \right.\right. \\ & \frac{(7 \ell^2-\ell m^2+10 \ell-9 m^2+3) r^2 \dot{\phi}^2}{2 (\ell+1) (2  \ell+3)} \nn \\ & \left.\left.  +\frac{(\ell^2+\ell+3) \dot{r}^2}{2 (2 \ell+3)} -\frac{ \ell (4\ell+11)  M}{2 (2 \ell+3) r(t)} +\ldots  \right] \right\}\;, \nn
\end{align}
\begin{align}
&\widehat{\cal A}^V_{\ell m} (t)  = -{\rm i} \eta  \mu \ell \dot{\phi} c_{\ell m}^1   r^{\ell+1} \left\{  1+\eta^2 \left[ \frac{(\ell^2+\ell+3) \dot{r}^2}{2 (2 \ell+3)} \right. \right. \\ \nn & -\frac{(\ell+1) (4\ell^2+3
   \ell-12) M}{2 \ell (2 \ell+3) r(t)} -\frac{ {\rm i} ( \ell-1) (\ell+3) m r   \dot{r} \dot{\phi} }{(\ell+2) (2 \ell+3)}\\ \nn & \left.\left. +\frac{(3 \ell^2-\ell m^2+8 \ell-4 m^2+4) r^2 \dot{\phi}^2}{2 (\ell+2) (2 \ell+3)}\right]+\ldots \right\}\;.
 \end{align}
  The analysis can be easily extended to higher PN orders if one keeps only the terms in  ${\cal A}^{\rm inst}_{\ell m}$.
    Plugging (\ref{almtime}) into (\ref{fourier}) and  performing the $t$ and $\omega$ integrals one finds
 \be
 W^{\rm inst}_{\ell m} (u)= \left( -{d\over du}\right)^\ell   \left[ e^{-im\phi(u) } \widehat{\cal A}^{\rm inst}_{\ell m}( u)  \right]\;.  \label{fourier2}
\ee
We will later compare these results against those available in literature for $\ell=2$.

\subsection{Source multipole moments }
\label{sect:multmom}

In this section, we compare the results (\ref{almtime}) against those obtained by the standard PM multipole (MPM) techniques.
  The latter often makes use of the harmonic radial coordinate  $r_H=r -M$ and the retarded time $u=T-R^*$. In this framework, the  instantaneous  part of the waveform  is written in terms of the source multipoles $I_{i_1,\ldots i_\ell}(u)$ and $J_{i_1 \ldots i_\ell}(u)$ evaluated along the trajectory. At leading order in the small mass ratio limit, one finds \cite{Blanchet:2013haa}
  \bea
  &&h^{\rm inst}(X) \underset{R\to \infty}{\approx} {4\over R} \sum_{\ell=2}^\infty  {1\over \ell!} \Big[  I^{(\ell)}_{i_1,\ldots i_\ell}(u)  \nn\\
 && -{2 \ell\over \ell+1}   {\bf n}_{j_1} \epsilon_{j_1  j_2 i_1 }  J^{(\ell)}_{j_2 i_2 \ldots i_\ell}(u)     \Big]{\bf e}_{i_1} {\bf e}_{i_2} {\bf n}_{i_3} \ldots {\bf n}_{i_\ell}\;,
  \eea
  where the superscript $(\ell)$ stands for the $\ell^{\rm th}$-time derivative, $ I^{(\ell)}= {d^\ell I\over du^\ell} $. The harmonic expansion coefficients $W_{\ell m}$ are defined as\footnote{ Here $\int d\Omega=\int_0^\pi d\Theta \,\sin\Theta \int_{0}^{2\pi} d\Phi$.}
 \bea\label{eq:W-ft}
 W_{\ell m} (u) &=& \int {d\omega\over 2 \pi} \, e^{-{\rm i}\omega u} W_{\ell m}(\omega)\nn\\  &=& \lim_{R\to \infty} {R\over 4G}  \int d\Omega\,  Y^{\ell m *}_{-2}(\Theta,\Phi)  \,  h(X)
 \eea
 leading to
 \be
 W^{\rm inst}_{\ell m}(u)= {1 \over \ell !}  {d^\ell \over du^\ell} \left[ I_{\ell m}(u)  -{2 \ell\over \ell+1}   J_{\ell m}(u)     \right]
 \label{hinstij}
 \ee
 with
 \begin{equation}
 \begin{aligned}
 I_{\ell m} &= \alpha_{\ell m}^{i_1 i_2; i_3 \ldots i_\ell}  I_{i_1,\ldots i_\ell}\;,\\  J_{\ell m}& = \alpha_{\ell+1 m}^{i_1 i_2;j_1 i_3 \ldots i_\ell}  \epsilon_{j_1 j_2 i_1 }  J_{j_2 i_2 i_3\ldots i_\ell}
 \end{aligned}
 \end{equation}
 and
   \be
\alpha^{\ell m}_{i_1 i_2; i_3 \ldots i_p}  = \int d\Omega\,  Y^{\ell m *}_{-2}(\Theta,\Phi)  \,   {\bf e}_{i_1} {\bf e}_{i_2} {\bf n}_{i_3} \ldots {\bf n}_{i_p}\;.
  \ee
The PN expansion of the source multipoles for some low values of $\ell$ has been computed in \cite{Mishra:2015bqa}. Specifying to $\ell=2$, one finds up to 1PN order
\begin{equation}
  \begin{aligned}
 I_{ij} (t) &= \mu x_{\langle i} x_{j \rangle} +\eta^2 \mu  \Big[ \Big( \frac{29  \dot{x}^2 }{42} -{5M\over  7 r_H}\Big) x_{\langle i} x_{j \rangle}  \\
  &- \frac{4 r \dot{r} }{7 }x_{\langle i} \dot{x}_{j \rangle} +\frac{11 r_H^2 }{21 } \dot{x}_{\langle i} \dot{x}_{j \rangle}   \Big]  +\ldots\;, \\
  J_{ij} (t) &= - \mu  x_{k_1} \dot{x}_{k_2} \epsilon_{k_1 k_2 \langle i} \Big[  x_{j\rangle} + \eta^2 \frac{5 \dot{r} r }{28} \dot{x}_{j\rangle} \\&  + \eta^2 x_{j\rangle}\Big(  \frac{13 \dot{x}^2 }{28} +{27 M\over 14 r_H}  \Big)
    \Big]+\ldots\;,
  \end{aligned}
\end{equation}
where the angle brackets on the indices indicate the symmetric traceless combination. Projecting on the basis of spin weighted spherical harmonics one finds
 \begin{align}
  I_{2,\pm 2} &= \ft12\sqrt{\ft{\pi}{5}} r_H^2 e^{\mp 2{\rm i} \phi}  \Big( 1+ \eta^2 \Big[  -{5M \over  7 r_H } +  { 9  \dot{r}_H^2\over 14 } \nn\\ & \mp   {\rm i} {10 r_H \dot{r}_H \dot{\phi} \over 21} +{ r_H^2 \dot{\phi}^2   \over 6} \Big] +\ldots \Big)\;,  \nn\\
   I_{2,0} &= -\sqrt{\ft{\pi}{30}} r_H^2 \left( 1+ \eta^2 \left[ -{5M \over  7 r_H }  + { 9  \dot{r}_H^2\over 14 } \right.\right. \nn \\ & \left.\left.  +{ 17 r_H^2 \dot{\phi}^2   \over 14} \right] +\ldots \right)\;,  \label{iij}\\
     J_{2,\pm 1} &= \ft12\sqrt{\ft{\pi}{5}} \mu  r_H^3  \dot{\phi}(t)  e^{\mp {\rm i} \phi}  \Big( 1+ \eta^2 \Big[ {27M \over  14 r_H }  + { 9  \dot{r}_H^2\over 14 } \nn\\&  \mp   {\rm i} {5  r_H \dot{r}_H \dot{\phi} \over 28} +{ 13 r_H^2 \dot{\phi}^2   \over 28} +\ldots   \Big] \Big)\;. \nn
 \end{align}
   Comparing against (\ref{almtime}) one finds  a perfect match after the identifications\footnote{The near-zone multipoles $I$ and $J$ are gauge-dependent quantities, so any comparison has to be performed by using the same coordinates system.} $r_H=r -M$ and
   \begin{equation}
\begin{aligned}
  e^{-{\rm i} m \phi} \widehat{\cal A}^U_{\ell m}(u) &= {(-1)^\ell\over \ell !}  I_{\ell m} (u)+O(\eta^3)\;,  \\
  e^{-{\rm i} m \phi}  \widehat{\cal A}^V_{\ell m}(u) &= {(-1)^{\ell+1}\over \ell !}  {2{\rm i} \ell  \over \ell+1} \, J_{2m} (u)+O(\eta^3)\;.
 \end{aligned}
   \end{equation}
following from (\ref{fourier2}) and (\ref{hinstij}).

\subsection{Tail and tail of tail contributions}
\label{sec:tails}

Tails are non-local in time ``hereditary" contributions to the wave form, originating from back scattering of multipolar waves off the Schwarzschild curvature  generated by the mass monopole $M$ of the source. They are represented as integrals over the past history of the source of multipole moments. The leading contributions to the $U$ part of the wave form are given by  Eq.~(196) of~\cite{Blanchet:2013haa}, which in the probe limit reads
\begin{align}
&W^{U,\rm tail}_{\ell m}(u)  = 2  M \int\limits_0^{\infty} d\tau \, I_{\ell m}^{(\ell+2)}(u-\tau)\Big[\ln \Big(\frac{c \tau}{2 b_0}\Big)+\kappa_{\ell}\Big] \nn  \\
& + 2  M^2 \int\limits_0^{\infty}  d \tau\, I_{\ell m}^{(\ell+3)}(u-\tau)\Big[\ln ^2\Big(\frac{c \tau}{2 b_0}\Big) \label{eq:196} \\
&+\Big(2 \kappa_{\ell}+{\beta_\ell^{[2]} \over 2} \Big) \ln (\frac{c \tau}{2 b_0})+\xi_{\ell} \Big]+\mathcal{O}(M^3) . \nn
\end{align}
In the frequency domain, we write
   \be
   I_{\ell m}^{\ell+n}(u-\tau)=\int {d\omega' \over 2\pi} e^{-{\rm i} \omega' (u-\tau)}(-{\rm i} \omega )^n\,    I^{\rm \ell}_{\ell m}(\omega)
   \ee
  leading to
  \begin{align}\label{eq:1962}
& W^{U,\rm tail}_{\ell m}(\omega)  = \int du\,  e^{{\rm i}\omega u} \,  W^{\rm U, tail}_{\ell m}(u) =  I^{\rm \ell}_{\ell m}(\omega) \Big[  2   \, {\rm i} \omega M\, (\ln(\omega) \nn\\
&-\kappa_{\ell}{-}\frac{{\rm i}\pi}{2}  {+} \gamma)  +  \omega^2 M^2( -\ln^2(\omega) + (\beta_\ell^{[2]}+4 \kappa_{\ell}+2{\rm i}\pi -4\gamma)  \nn \\
&    \ln(\omega) -2\tilde \xi_{\ell}) +\mathcal{O}(M^3) \Big] \;,
\end{align}
 where the $u$-integral produces a delta-function that allows the $\omega'$-integration and $\tau$-integrals to be computed  with the help of the master integral
  \begin{align}
  \int\limits_0^\infty d\tau\, e^{{\rm i} \omega \tau} \tau^a&={\Gamma(a+1)\over (-{\rm i} \omega )^{a+1} } ={{\rm i}\over \omega} \Big( 1{-} a
  [ \gamma{+}\ln({-}{\rm i} \omega)]\nn\\
  &{+} {a^2\over 2} \Big[ (\gamma{+}\ln({-}{\rm i} \omega))^2 {+}\ft{\pi^2}{6} \Big]{+}\ldots\Big) \;.
  \end{align}
  The results (\ref{eq:1962}) can be easily matched against those obtained from perturbation theory.  Indeed (\ref{eq:1962}) can be written as
  \begin{align}\label{eq:19621}
& W^{U,\rm tail}_{\ell m}(\omega)   =- I^{\rm \ell}_{\ell m}(\omega) +  I^{\rm \ell}_{\ell m}(\omega) \left[ 1-2   \, {\rm i} \omega M\, (\kappa_\ell+\gamma)+\ldots\right]\nn\\
& \left[  1 +{\cal B}_\ell^{\rm tail}(\omega,r)   \right] \; ,
\end{align}
 where the $I^{\rm \ell}_{\ell m}$-term   removes the instantaneous contribution and
  \begin{align}\label{atail}
   &  {\cal B}^{\rm tail}_{\ell}(\omega,r)  =  (-2{\rm i} \omega r )^{  \widehat{\ell}-\ell}  ({-}4{\rm i} \omega M)^{ 2{\rm i} \omega M}-1
      \\ &  =2{\rm i} \omega M \log  ({-}4{\rm i} \omega M)+
  \omega^2 M^2\nn\\
  & \left[\beta_\ell^{[2]} \log(-2{\rm i} \omega r )  -2 \log^2  ({-}4{\rm i} \omega M) \right] +\ldots \nn
  \end{align}
     collects all $\log \omega$ terms in ${\cal A}^U_{\ell m}(\omega)$, that originates from  $\mathfrak{R}^{\rm flow}_{\ell}$ and the first factor in $\mathfrak{R}^{\rm tail}_{\ell}$.
   Indeed the first line of~\eqref{eq:196} reproduces the 1.5PN contributions  of~\eqref{aulm} with the log term originating from $\mathfrak{R}^{\rm tail}_{\ell}$. On the other hand,
   the log terms in the second line of~\eqref{eq:196} reproduces the combination (\ref{atail}) arising at order 3PN from the expansion of $\mathfrak{R}^{\rm tail}_{\ell}$ and $\mathfrak{R}^{\rm flow}_{\ell}$.
    The coefficient  $\beta_\ell^{[2]}$ was introduced in~\eqref{lhatl} and represents the tail of tail contributions discussed in~\cite{Blanchet:1987wq}, see in particular (A6) of that reference. The same result was obtained in~\cite{Almeida:2021jyt} in the context of the Effective Field Theory (EFT) approach~\cite{Goldberger:2009qd} and more recently from the amplitudes approach in~\cite{Georgoudis:2023eke,Georgoudis:2024pdz,Bini:2024rsy}.

  A completely analogous discussion holds for the current-type $V$ contributions with $I$ replaced by $J$ and $\kappa_\ell$ by $\pi_\ell$.  Tail contributions originate from the same function
  ${\cal B}^{\rm tail}_{\ell}(\omega,r)$ given by (\ref{atail}).In particular the coefficients $\beta_\ell^{[k]}$ introduced in~\eqref{lhatl} do not depend on $m$ and equally contribute to mass-type and current-type moments. This is in tension with the result of~\cite{Almeida:2021jyt}, where a different coefficient $\beta^{\rm mag}_\ell \neq \beta^{[2]}_\ell$
  was introduced to describe the   $\log\tau$ term at order 3PN of the current-type version of~\eqref{eq:196},  see also (198) and (200) of ~\cite{Blanchet:2013haa}\footnote{The first mismatch appears for $\ell=3$ where~\cite{Almeida:2021jyt} gives $\beta^{\rm mag}_3=-13318/10395$, while~\eqref{beta2def} yields $\beta^{[2]}_3=-26/21$.}.
  We have verified our results for $\beta_\ell^{[2]}$ and $\beta_\ell^{[4]}$ by matching the $\ln(v)$-dependent contributions at order $6$PN with the formulae provided in ~\cite{Tanaka:1996lfd,Fujita:2012cm} for the energy flux generated by circular motion. Our results have also been confirmed by an independent computation  of the tail-of-tail contribution for the current octupole moment by Blanchet and Faye\footnote{We would like to thank L.~Blanchet and G.~Faye for performing this check and letting us know about their result.} using the MPM approach.
We remark that only $\mathfrak{R}^{\rm flow}_{\ell}$ contributes to the energy flux since $\log \omega$-terms in $\mathfrak{R}^{\rm tail}_{\ell}$ resum to a pure phase $\omega^{2{\rm i} \omega M}$.

\section{Hyperbolic encounters}
\label{sec:hyp}

 In this section, we compute the wave form radiated by the hyperbolic scattering of a light particle in the Schwarzschild geometry.
 The results will be compared against those recently obtained using MPM and amplitude techniques.
  We adopt a near Keplerian ansatz of the trajectory given as
 \bea
r_H(v) & = &  M a_r   ( e_r \cosh v- 1 )\;,   \nn\\
t(v) &= &    M a_r   \gamma_t \left[   e_r \sinh v-\gamma_v\, v   +f( \phi_0  )  \right]\;, \nn\\
\phi(v) & = & \gamma_\phi  \, \phi_0(v)  + g(\phi_0 )\;,  \label{geodesics2a}
\eea
Here  $a_r,e_r,\gamma_t,\gamma_v,\gamma_\phi$ are constants ($v$-independent) parameterizing the geodesics (asymptotic velocities, impact parameter, eccentricities, etc.) which can be computed in a PN expansion (See for instance~\cite{Cho:2018upo} for an explicit 3PN expression). Furthermore
\be
\phi_0(v) =2   {\rm Arctan} \left[ \sqrt{e_\phi+1\over e_\phi -1}  \tanh(\ft{v}{2}) \right]\;.  \nn\\
\ee
and  $f(\phi_0)$, $g(\phi_0)$ are  functions of $\phi_0$.  . All variables and functions  entering in the ansatz (\ref{geodesics2a}) are determined  in terms of the energy $E$ and the angular momentum $J$ by the geodesic equations  (\ref{uvel}) that can be equivalently written as
\bea
{ \phi'(v) \over t'(v) } &=& { f(r) J  \over E \, r^2}\;,    \nn\\
{ r'(v)^2 \over t'(v)^2 } &=&   {f(r)^2\over E^2}\left[ E^2-f(r) \left(\mu^2+{J^2 \over r^2} \right) \right]\;. \label{geodesics2b}
\eea
  These equations determine all the parameters and functions entering in the ansatz  (\ref{geodesics2a}) as a function of two independent variables that can be taken to be
   the energy and the angular momentum $(E,J)$ or, as we will do here, the pair $(a_r,e_r)$ parameterizing  the velocity and eccentricity of the orbit (see below).

 \subsection*{PN expansion}

  The geodesics equation (\ref{geodesics2b}) can be solved perturbatively in the limit
  $a_r \to \infty$, keeping $e_r$ exact. One finds
\begin{align}
\gamma_t &= \sqrt{ a_r-1 },  \gamma_v=    1-4 a_r^{-1}  ,  e_\phi =  e_r  {+}\ldots, \gamma_\phi = 1{+}\ldots, \nn  \\
  E &= \mu \sqrt{a_r-1\over a_r-2}  , f(\phi_0 )  =  0+\ldots  , g(\phi_0 )  =  0+\ldots , \nn\\
  J &=  M \mu \sqrt{ a_r(e_r^2-1)} \left( 1+{  e_r^2+2\over a_r(e_r^2-1)} +\ldots \right),
 \label{ks1}
\end{align}
 where the dots stand for corrections of order $a_r^{-2}$. By sending $v\to \infty$ in (\ref{geodesics2a}), it is easy to see that in the limit $a_r \to \infty$, the
 particle velocity  $\gamma_t^{-1}\sim a_r^{-1/2}$ vanishes, so the limit $a_r \to \infty$ corresponds  to the PN  expansion of the geodesics.

   \subsection*{PM expansion}

   Alternatively, equations (\ref{geodesics2b}) can be solved order by order in $1/e_r$, in the limit  where $e_r \to \infty$, $M\to 0$, keeping finite their product $M e_r$  and $a_r$.
 This corresponds therefore to the PM limit. To leading order one finds
\begin{align}
\gamma_t &=  \sqrt{ a_r-1 } , \quad \gamma_v=    1-4 a_r^{-1}   ,  \quad \frac{e_\phi}{e_r} =  1 +\ldots\;,  \nn\\
 E &=  \mu \sqrt{a_r-1\over a_r-2}, \qquad   \frac{J}{e_r} =  { \mu M a_r \over \sqrt{a_r-2} } +\ldots \;,
 \label{ks2} \\
 f(\phi_0 ) & =  0+\ldots  , \quad g(\phi_0 )  =  0+\ldots  , \quad    \gamma_\phi =  1 +\ldots \;, \nn
  \end{align}
  where the dots stand for corrections of order $e_r^{-2}$.

  \subsection*{The wave form}

The wave form is computed by the integral along the trajectory
\be
 W_{\ell m} (\omega)=    \int \! dv \, t'(v) \, e^{-im\phi(v)+i\omega t(v) } {\cal A}_{\ell m}(\omega, r(v)) \;. \label{hlmhyp}
\ee
The integral can be explicitly performed in the two extreme cases describing the PN and PM limits, where the geodesics are given in  (\ref{ks1}), (\ref{ks2}).
 The main simplification in these limits comes from the fact that $f,g\to 0$, $\gamma_\phi\to 1$ and therefore the trajectories (\ref{geodesics2a}) can be written in the simple form
 \bea
{\rm i} \omega r_H(v) & \approx &    {\rm i}  u \cosh v\,  \gamma_t^{-1}  -     \alpha\,  (\gamma_t \gamma_v)^{-1} \;,  \nn\\
{\rm i} \omega t(v) &\approx  &  {\rm i}  u \sinh v-\alpha \, v \;,   \nn\\
{\rm i} \omega r_H(v) e^{-{\rm i}  \phi_0(v)} & \approx &   \gamma_t^{-1}  \Big[ \sqrt{u^2+\alpha^2 \gamma_v^{-2} } \sinh v  \label{geodesics2c} \\
&&+({\rm i}  u - \alpha   \cosh v \gamma_v^{-1}  )  \Big] \;,   \nn
\eea
with variables
 \be
 u=\omega  M a_r   \gamma_t   e_r    , \quad \alpha = {\rm i} \omega  M a_r  \gamma_t \gamma_v
 \ee
 parameterising the wave frequency and the orbit eccentricity
 \be
\omega = -\frac{ {\rm i} \alpha }{M a_r \gamma _t \gamma _v} \qquad , \qquad e_r= \frac{i u \gamma _v}{\alpha }\;.
 \ee
Using (\ref{geodesics2c}) and some trigonometric identities, one can write the integrand in (\ref{hlmhyp}) in the form
\be
 t'(v) e^{-im\phi(v)  } {\cal A}_{\ell m}(\omega, r(v)) \!=\! \sum_{n=0}^\infty \sum_{\nu=0}^1 c_{n,\nu} (\sinh v)^n  (\cosh v)^{\nu},
\ee
with $c_{n,\nu}$ some coefficients.
Plugging this into (\ref{hlmhyp}) one finds
\be
W_{\ell m} (\omega)= \sum_{n=0}^\infty \sum_{\nu=0}^1 c_{n,\nu} \,{\cal I}_{n,\nu}\;,
\ee
 with ${\cal I}_{n,\nu}$ the master integrals
 \bea
&&{\cal I}_{n,\nu} =  \int_{-\infty}^\infty    dv  e^{ {\rm i} u \sinh v-\alpha v } \,  (\sinh v)^n  (\cosh v)^{\nu}\nn\\
&&=  (-{\rm i} \partial_u)^n  \left[ 2e^{-{{\rm i}\pi \alpha\over 2}}   \left( {   \alpha \over  {\rm i} u} \right)^{ \nu}
 K_\alpha (u) \right]   \label{integrals}\;.
 \eea
    The final result can be expressed in terms of the modified Bessel functions $K_\alpha$ and $K_{\alpha+1}$ with the help of the identities
  \bea
   {dK_{\alpha}(u) \over du}  &=&  {\alpha \over u} K_{\alpha}(u){-}K_{\alpha+1}(u)   ,\nn\\  K_{\alpha{+}1}(u)& =& K_{\alpha{-}1}(u){+}{2 \alpha  \over u} K_{\alpha}(u)\;.
  \eea
 In the rest of this section we display some explicit formulae for the first few leading terms of the wave form in the PN expansion and compare them against results
 obtained via the MPM and amplitude techniques.

\subsection*{Examples: $W^{0PN}_{2 m}$   }

We consider first the PN expansion, restoring the light velocity, and sending  $\eta \to 0$ and  $a_r \to \infty$.
At leading order
\be
 {\cal A}^{\rm U}_{\ell m}   \approx  \eta^\ell\, E\, c_{\ell m}^0\,    ( {\rm i} \omega r )^{\ell} , \quad
 {\cal A}^{\rm V}_{\ell m}   \approx  \eta^{\ell+1}\,  J\,\ell \, c_{\ell m}^1\, \omega    ( {\rm i} \omega r  )^{\ell-1}
 \ee
 and $\gamma_v \to 1$, $\gamma_t\to \sqrt{a_r}$. The leading contribution to the wave form comes then from
    $\ell=2$ and using (\ref{geodesics2c}-\ref{integrals}) can be written as
 \bea
 &&W^{0PN}_{20} = {-}2 \sqrt{\frac{2 \pi a_r }{15}} e^{-\frac{1}{2} i \pi  \alpha }  \mu  M   K_{\alpha }(u)\;, \nn\\
&&W^{0PN}_{2,\pm 1} =  {-} \frac{2 {\rm i} }{3} \sqrt{\frac{\pi }{5}} e^{-\frac{1}{2} i \pi  \alpha } {\mu  M\over u^2} (2u^2+\alpha^2) \nn\\
&&\left( \mp  \sqrt{\alpha ^2+u^2}
   K_{\alpha }(u)+\alpha  K_{\alpha }(u)-u K_{\alpha +1}(u)\right)\;, \nn\\
&& W^{0PN}_{2,\pm 2} = {-} 2 \sqrt{\frac{\pi a_r }{5}} e^{-\frac{1}{2} i \pi  \alpha } \frac{ \mu  M }{u^2}
 \Big[ 2 u K_{\alpha +1}(u) \nn\\
 &&(\alpha ^2+u^2\pm  \sqrt{\alpha
   ^2+u^2})  +K_{\alpha }(u) (2
   (1-\alpha ) \alpha ^2\nn\\
   &&+(1-2 \alpha) u^2 \pm 2  ((\alpha -1) \alpha +u^2) \sqrt{\alpha ^2+u^2})\Big]\;. \nn
\eea
In this limit $ u \approx \omega  M a_r^{3/2}     e_r $ and   $ \alpha  \approx {\rm i} \omega  M a_r^{3/ 2}   $.

\subsection*{Examples: $W^{1PN, G^1}_{2 m}$   }

 Next we consider the 1PN expansion of the wave form at leading order in the limit $e_r \to \infty$, i.e. $\alpha \to 0$.
 Proceeding as before and using
  \bea
  {dK_{\alpha}(u) \over d \alpha}\Big|_{\alpha=0}  = 0 ,\quad  {dK_{\alpha+1}(u) \over d \alpha}\Big|_{\alpha=0}  = {K_0\over u}\;,
  \eea
 one finds
  \bea
 &&W^{\rm 1PN, G^1}_{20}   \underset{\alpha \to 0}{\approx}  {-} 2 \sqrt{\frac{2 \pi a_r  }{15}} \mu  M  \Big( -\frac{20
   K_0(u)+26 u K_1(u)}{7  a_r}\nn\\
   &&+K_0(u)\Big)\;, \nn\\
  &&W^{\rm 1PN, G^1}_{2,\pm2}  \underset{\alpha \to 0}{\approx}   {-}  2 \sqrt{\frac{\pi  a_r }{5}} \mu  M  \Big[ K_0(u) ( \pm 2 u
   +1) \nn\\
   &&  +2K_1(u) (  u
   \pm 1)+ \frac{1}{21 a_r} (
   K_0(u) (2 u^2\mp 41 u -60+K_1(u)\nn\\&& ( \pm  u^2
    -20 u\pm 24  ))+\ldots\Big]\;, \nn\\
 && W^{\rm 1PN, G^1}_{2,\pm1}  \underset{\alpha \to 0}{\approx}   {+} \ft{4 {\rm i} }{3}
   \sqrt{\ft{\pi }{5}} \eta  \mu  M u \Big[  K_0(u) \pm
   K_1(u)\nn\\
  && +    \frac{ 1}{2 a_r}   ( 5 K_0(u) \mp 3 K_1(u))  \Big]\;.
    \eea
The results match (3.18), (3.19), (3.25) in \cite{Bini:2024rsy} and formulae in \cite{DiVecchia:2023frv} after the identifications  ${ W_{\ell m}^U= U_{\ell m} } $, $ W_{\ell m}^V= V_{\ell m} $ and
  \be
 a_r=\vi^{-2}+2 \qquad ,\ \qquad  u=u_D (1+\ft{p_\infty^2}{2}+\ldots)\;,
   \ee
   where  $u_D=\omega b/p_\infty$ is the u-parameter in the notation of \cite{Bini:2024rsy} and $p_\infty=p/\mu$.

\section{ Soft limit }
\label{sec:soft}

In this section we derive an  exact formula (all orders in $G$ and $p_\infty$) for the universal part of the gravitational wave form in the probe and soft limit $\omega \to 0$.

\subsection{The Weinberg theorem}

The gravitational wave form emitted by a particle moving in a Schwarzschild geometry can be computed by a five point scattering amplitude involving two incoming and two outgoing massive particles, let us say of masses $\mu$, $M$, and an outgoing graviton in flat space. The wave form represents now a perturbation $ g_{\mu\nu}=\eta_{\mu\nu}+h_{\mu\nu}$ of the flat metric $\eta_{\mu\nu}$.
We are interested on the limit where the graviton is soft, i.e. it carries away a tiny fraction of the total momentum.
The leading term in the soft limit of the gravitational wave produced by the scattering of $N$ particles in flat space is known to be universal (this is the content of the Weinberg theorem) and given by \cite{Weinberg:1965nx}
\be \label{eq:nW}
  h_{\mu\nu}(\omega,\mathbf{n}) =\int dt\, e^{i\omega t} h_{\mu\nu}(X) \underset{\omega \to 0}{\approx} - \frac{4 G}{R}  \sum_{j=1}^N {\rm i} {p_j^\mu p_j^\nu\over p_j \cdot k}\;,
\ee
where $p_i$ are the external ``hard'' momenta and
\be
k^\mu=\omega n^\mu=\omega (1,\bf{n})
\ee
is the momentum of the emitted soft graviton. The vector ${\bf n}$ was introduced in (\ref{unitvector}) and similarly we define
  \begin{equation}
    \label{eq:enobold}
    e^\mu = (0,\bf{e})\;
  \end{equation}
  in terms of the polarization vector ${\bf e}$ introduced in (\ref{polarizationvector}). For $2\to 2$ scattering one finds
\be
h_{\rm soft}  \equiv h^{\rm soft}_{\mu\nu}  e^\mu  e^\nu =- \frac{4 G}{R} \sum_j {\rm i} { (p_j \cdot e )^2 \over  p_j \cdot k} \equiv \frac{4 G}{R} W_{\rm soft}\;,
\ee
with $j=(1_{\rm in}, 2_{\rm in},1_{\rm out},2_{\rm out})$ running over all outgoing momenta
\begin{equation}\begin{aligned}
p^\mu_{1_{\rm in}} &=(-E_1,-p \sin\ft{\Delta\phi}{2}, -p \cos\ft{\Delta\phi}{2},0) \;, \nn\\
  p^\mu_{2_{\rm in}} & =(-E_2,p \sin\ft{\Delta\phi}{2}, p \cos\ft{\Delta\phi}{2},0) \;, \nn \\
p^\mu_{1_{\rm out}} & =(E_1,-p \sin\ft{\Delta\phi}{2}, p \cos\ft{\Delta\phi}{2},0)  \;,\nn\\
p^\mu_{2_{\rm out}} &= (E_2,p \sin\ft{\Delta\phi}{2}, -p \cos\ft{\Delta\phi}{2},0)\;,
\end{aligned}\end{equation}
with $\Delta\phi$ the deflection angle and
\be
E_1 \simeq E=  \sqrt{\mu^2+p^2}\;.
 \label{E1E2}
\ee
In this regime the contributions involving the heavy particles are suppressed and one finds
\bea
&&W_{\rm soft} \underset{\mu \ll M}{\approx} - {\rm i} \sum_{j=1_{\rm in}, 1_{\rm out} } { (p_j \cdot e )^2 \over p_j \cdot k}\nn\\
&&=  {  {\rm i} p^2 \over 2 \omega  }  \sum_{\epsilon=\pm}  \epsilon { \left[ \cos(\Phi+\frac{\epsilon \Delta \phi}{2})+{\rm i} \cos\Theta \sin(\Phi+\frac{\epsilon \Delta \phi}{2}) \right]^2 \over E-  p \sin\Theta \sin(\Phi+\frac{\epsilon \Delta \phi}{2})}\;. \label{weinberg}
\eea

We notice that the soft wave form  is universal {\it i.e.} it depends only on the initial and final momenta  with no reference to the details of the scattering amplitude.
As already mentioned in the Introduction, in \cite{Saha:2019tub}  it was conjectured that this universality property is shared by an infinite tower of logarithmic terms growing as $\omega^{n-1} (\log \omega)^{n} $. Explicit formulae were
found for the first two terms $n=1,2$ in \cite{Sahoo:2018lxl,Sahoo:2020ryf}.  Here we confirm this expectation and use perturbation theory to derive an exact formula (both in $G$ and $p$), in the probe limit, for the soft wave form, where the entire tower of logarithmic terms $\omega^{n-1} (\log \omega)^{n} $ is resummed into an exponential.

\subsection{The soft wave form from perturbation theory} \label{ssec:leadlogsof}

  In the limit where $\omega \to 0$, the integral (\ref{za}) is dominated by the region where $t$ is large,  i.e.  $t=\pm \infty$.
   In this limit, the geodesic  can be approximated by two straight lines  deflected by an angle $\Delta \phi$.  The geodesic equations are solved by the ansatz
   \bea
t(r) & \underset{r\to \infty}{\approx} &  \pm   ( \gamma_t   r  -  M C \ln r +\ldots )\;, \nn\\
\phi(r) & \underset{r\to \infty}{\approx} &    \pm \left(  {\pi \over 2} +  {\Delta \phi\over 2}   \right)  +\ldots\;. \label{geodesics3}
\eea
Imposing the geodesic equations (\ref{eq:dert})
\be
\begin{gathered}
{ \phi'(r) \over t'(r) } = { f(r) J  \over E \, r^2}\; ,\\
{ 1\over t'(r)^2 } =  {f(r)^2\over E^2}\left[ E^2-f(r) \left(\mu^2+{J^2 \over r^2} \right) \right] \label{geodesics2d}
\end{gathered}
\ee
 order by order in the limit $1/r$ one finds
 \be
 \begin{gathered}
E= {\mu \gamma_t \over \sqrt{\gamma_t^2-1} } \equiv \mu \sigma \;,\\   C=  \gamma_t  (\gamma_t^2-3) = \frac{\sigma (3-2\sigma^2)}{(\sigma^2-1)^{3/2}}\;.
\label{energygammat}
 \end{gathered}
\ee
 In addition, since $r\gg M$  the only non-trivial contribution to ${\cal A}_{\rm \ell m}$  comes from the
 ${\cal L}^0$ part of  ${\cal A}^U_{\ell m}$ leading to
  \bea
 {\cal A}_{\ell m} \underset{r\to \infty}{\approx}   E    \,(-4{\rm i} \omega)^{2 {\rm i} \omega M}   \,c_{\ell m}^0\,  \left(1+ \dot{r} \right)^2  ( {\rm i} \omega r)^{\ell} e^{- {\rm i} \omega r  } \nn\\ _1F_1(\ell+3;2 \ell+2; 2{\rm i} \omega r)\;.  \label{almsoft}
 \eea
Expanding the right hand side of (\ref{almsoft}) in powers of ${\rm i} \omega r$, and writing
 \be
e^{- {\rm i} \omega r  } \, _1F_1(\ell+3;2 \ell+2; 2{\rm i} \omega r) \equiv \sum_{n=0}^\infty \mathfrak{c}_{n} ( {\rm i} \omega r)^{n}\;,
\ee
the integral (\ref{za}) can be written as
\bea
&&W_{\ell m}  \underset{\omega \to 0}{\approx}    E    \gamma_t   \,\omega^{2 {\rm i} \omega M}   \,c_{\ell m}^0\,   \sum_{n=0}^\infty \sum_{\epsilon=\pm}  \mathfrak{c}_{n}  e^{  - {{\rm i} m \epsilon\over 2}
(\pi+\Delta \phi ) } \nn\\
&&  \left(1+  \epsilon \gamma_t^{-1} \right)^2  \int\limits_{0}^\infty  dr \, e^{{\rm i} \epsilon \omega (\gamma_t  r - CM \ln r)}   ( {\rm i} \omega r)^{\ell+n      }, \label{wlm1}
\eea
where we have split the domain of integration into two regions labelled by $\epsilon=\pm$, where $t\approx \epsilon (\gamma_t r-MC \ln r)$ and $\dot{r} \approx \epsilon\gamma_t^{-1} $. Here and in the following we keep only the
 leading logarithmic terms, i.e. those contributing  as $\omega^{n-1} \ln^n \omega$ for $n=0,1,\ldots$, in the small frequency limit. Indeed, at this order,  the integral on $r$ can be extended all the way down to $r=0$ leading to\footnote{As usual we use the $i\epsilon$ prescription to make the integral over $r$ convergent mapping it to the integral representation of the Euler Gamma function
\be
\int_0^\infty dx \,e^{-\alpha x} x^\beta=\alpha^{-\beta-1} \Gamma(\beta+1)
\ee
Notice that the precise values of the lower extreme of integration is immaterial for the leading-log terms we are interested in.
}
\bea
&&({\rm i} \omega)^{\ell+n}  \int_{0}^\infty  dr \,   e^{{\rm i} \epsilon \omega \gamma_t  r}    r^{\ell+n  - {\rm i} \epsilon \omega C M   } \nn\\
&&\underset{\omega\to 0}{\approx} (\ell+n)!  (-  \epsilon  \gamma_t  )^{-\ell-n   -1}  ({\rm i}  \omega  )^{  {\rm i} \epsilon \omega C M-1}
\eea
Plugging this into (\ref{wlm1}) one finds
\bea
W_{\ell m} && \underset{\omega \to 0}\approx       \frac{E\gamma_t}{{\rm i} \omega}   \,\omega^{2 {\rm i} \omega M}   \,c_{\ell m}^0\,   \sum_{n=0}^\infty \sum_{\epsilon=\pm}  \mathfrak{c}_{n}  e^{ - {{\rm i} m \epsilon\over 2}
(\pi+\Delta \phi ) }  \nn\\&& \left(1+ \epsilon \gamma_t^{-1} \right)^2  (\ell+n)!  ( - \epsilon  \gamma_t  )^{-\ell-n   -1}  \omega^{  {\rm i} \epsilon \omega C M} \;. \label{wlm2}
\eea
 The sum over $n$ can be performed using the identity
     \bea
     &&\sum_{n=0}^\infty    \mathfrak{c}_{n} (n{+}\ell)!  (-x)^{n}
   \left(1+  x\right)^2 \nn\\
   &&= \ell! \,  {}_2F_1 ( \ft{\ell-1}{2}, \ft{\ell}{2},\ell+\ft32, x^{2} )
     \eea
 that can be checked expanding around $x=0$.  Plugging this into  (\ref{wlm2}) one finds
\bea
W_{\ell m}  &\underset{\omega \to 0}{\approx}&      \frac{{\rm i}^m \,E}{\omega \gamma_t^\ell}   \,\omega^{2 {\rm i} \omega M}   \,c_{\ell m}^0\,   \ell! \,  {}_2F_1 ( \ft{\ell-1}{2}, \ft{\ell}{2},\ell+\ft32, \gamma_t^{-2} )\nn\\
&&  \sum_{\epsilon=\pm}  ( {\rm i}  \epsilon) e^{ - {{\rm i} m \epsilon  \Delta \phi  \over 2} }       \omega^{  {\rm i} \epsilon \omega C M}   \label{wlm3}
\eea
or equivalently
\bea
W_{\ell m}  &&\underset{\omega \to 0}{\approx}   -2 { {\rm i}^m \,E    \over \omega \gamma_t^\ell }   \,\omega^{2 {\rm i} \omega M}   \,c_{\ell m}^0\,   \ell! \,  {}_2F_1 ( \ft{\ell-1}{2}, \ft{\ell}{2},\ell+\ft32, \gamma_t^{-2} )\nn\\
   && {\rm Im} \left[  e^{ - {{\rm i} m   \Delta \phi  \over 2} }       \omega^{  {\rm i}  \omega C M}   \right]\;.
\eea
To compare against (\ref{weinberg})  we have to sum over the whole tower of spherical harmonics
\be
 W_{\rm soft} (\omega,\Theta,\Phi)=  \sum\limits_{\ell,m} W_{\ell m}(\omega) Y^{\ell m}_{-2}(\Theta,\Phi)\;.
 \ee
 Remarkably, the sum can be explicitly performed using the identity
 \begin{gather}
    \sum_{\ell,m}  Y_{-2}^{\ell m}(\Theta,\Phi) { {\rm i}^m c_{\ell m}^0\,  \ell!\over \gamma_t^{\ell-2}}  {}_2F_1 ( \ft{\ell-1}{2}, \ft{\ell}{2},\ell+\ft32, \gamma_t^{-2})\nn\\ = -\frac{1}{2} { \left[ \cos(\Phi)+{\rm i} \cos\Theta \sin(\Phi) \right]^2 \over 1-  \gamma_t^{-1} \sin\Theta \sin(\Phi)}. \label{identity2}
    \end{gather}
   This equation can be checked order by order in an expansion for $\gamma_t^{-1}$ small. Plugging (\ref{identity2})  into (\ref{wlm3}) one finds
   \begin{widetext}
   \begin{equation} \label{hsoftfinal}
   \begin{aligned}
     W_{\rm soft}(\omega,\Theta,\Phi)  \underset{\omega \to 0}{\approx}    {  {\rm i}  \,E   \omega^{2 {\rm i} \omega M-1}   \over 2  \gamma_t^2  }   \,   \sum_{\epsilon=\pm}   \epsilon  \omega^{ - {\rm i} \epsilon \omega C M} %\\
     { \left[ \cos(\Phi+\frac{\epsilon \Delta \phi}{2})+{\rm i} \cos\Theta \sin(\Phi+\frac{\epsilon \Delta \phi}{2}) \right]^2 \over 1- \gamma_t^{-1} \sin\Theta \sin(\Phi+\frac{\epsilon \Delta \phi}{2})}\\   \underset{\omega \to 0}{\approx}  % \\
       - {\rm i}  \omega^{2 {\rm i} \omega M}   \left[   { (p_{\rm in}  \cdot e )^2 \over p_{\rm in} \cdot k}  \omega^{-{\rm i}\, \omega M C  } + { (p_{\rm out}  \cdot e )^2 \over p_{\rm out} \cdot k}  \omega^{{\rm i}\, \omega M C  }\right]\;,
   \end{aligned}
 \end{equation}
\end{widetext}
     where we used the relation  $\gamma_t=E/p$ that follows from (\ref{E1E2}) and (\ref{energygammat}). After performing the sum~\eqref{identity2} over the harmonics, one obtains the full soft waveform at arbitrary velocities with a smooth ultrarelativistic limit~\cite{DiVecchia:2022nna}.
     Expanding (\ref{hsoftfinal})  for small $\omega$, one reproduces (\ref{weinberg}) for the leading term diverging as $\omega^{-1}$. In addition (\ref{hsoftfinal}) resums an infinite tower of logarithmic corrections. The results at order  $\log \omega$ and  $\omega (\log \omega)^2$ match those coming from \cite{Sahoo:2018lxl,Sahoo:2020ryf} in the probe limit and
     they are in agreement with the probe limit of the recent conjecture of~\cite{Alessio:2024onn}, see for instance (2.75) of that paper which reads\footnote{We thank the authors of~\cite{Alessio:2024onn} for enlightening discussions on this point.}
     \begin{widetext}
       \begin{align}
         W_{\rm soft} & = - \frac{{\rm i}  \omega^{2 {\rm i} \omega G E_T}}{E_T \omega}   \left[\Big((p_{1{\rm in}} \cdot e)(p_{2{\rm in}}  \cdot n) -(p_{2{\rm in}} \cdot e)(p_{1{\rm in}}  \cdot n)\Big)^2 \frac{\omega^{-{\rm i}\, \omega E_T C  }}{(p_{1_{\rm in}} \cdot n)(p_{2_{\rm in}} \cdot n)} \right. \nn\\ &\left.
              - \Big((p_{1{\rm out}} \cdot e)(p_{2{\rm out}}  \cdot n) -(p_{2{\rm out}} \cdot e)(p_{1{\rm out}}  \cdot n)\Big)^2  \frac{\omega^{{\rm i}\, \omega E_T C  }}{(p_{1_{\rm out}} \cdot n)(p_{2_{\rm out}} \cdot n)} \right],
             \label{eq:271rew}\end{align}
    \end{widetext}
    where $E_T=E_1+E_2$. This result reduces to~\eqref{hsoftfinal} in the near probe limit, where $G E_T \approx M$ and $G p_2^\mu\approx (M,0)$. As a further evidence for the result~\eqref{eq:271rew} beyond the probe approximation, it is interesting to point out that this conjecture reproduces the soft limit of the full Newtonian waveform for the $2 \to 2$ scattering process~\cite{Alessio:2024onn}. Weinberg's formula~\eqref{eq:nW} for the leading $\omega^{-1}$ term of the soft waveform holds for a process with $N$-particles and~\cite{Alessio:2024onn} proposed a generalisation for all leading logarithmic terms $\omega^{n-1} \ln^n\omega$, see (2.76) of~\cite{Alessio:2024onn} and Sect.5.2 of~\cite{Sen:2024bax} for a further discussion. It is again straightforward to check their proposal in the probe limit where there are $N-1$ non-interacting particles of masses $\mu_i\ll M$ propagating in the Schwarzschild background. At leading (linear) order in each $\mu_i$ only the last term in (2.76) of~\cite{Alessio:2024onn} survives and reduces to a single sum over the light particles
   \begin{widetext}
      \begin{equation}
        \label{eq:hsoftN}
         W_{\rm soft}(\omega,\Theta,\Phi)  \underset{\omega \to 0}{\approx} - {\rm i}  \omega^{2 {\rm i} \omega M} \sum_{j=1}^N  \left[   { (p_{{\rm in}\,j}  \cdot e )^2 \over p_{{\rm in}\, j} \cdot k}  \omega^{-{\rm i}\, \omega M C(\sigma_j)  } + { (p_{{\rm out}\,j}  \cdot e )^2 \over p_{{\rm out}\,j} \cdot k}  \omega^{{\rm i}\, \omega M C(\sigma_j)  }\right]\;.
     \end{equation}
   \end{widetext}
As expected, this is exactly the sum of $N-1$ contributions, each one taking the form of~\eqref{hsoftfinal} with the momenta $p_j$ and the Lorentz factor $\sigma_j$ of the $j^{\rm th}$ light particle. It would of course be important to go beyond the leading order probe limit to test the other two terms of (2.76) of~\cite{Alessio:2024onn}.

\section{Summary and outlook}
\label{summary}

In this paper we applied black hole perturbation theory to the study of the gravitational wave forms produced by particles moving in the Schwarzschild geometry along an arbitrary (open or closed) trajectory. The recent interest in the study of open trajectories and unbounded binary systems has been triggered by the impressive development in the last few years of the computation of scattering amplitudes in gravity by means of quantum field theory techniques and the advent of a new space based generation of experiments that can bring binary system of increasingly large size inside the range of experimental detection.
Our results are obtained exploiting the correspondence between supersymmetric gauge theories and gravity that allows to compute the PN expansion of the wave form as a double series.
\\
Formulae (\ref{aulm}), (\ref{avlm}) and (\ref{almtime}) give the PN expansion of the gravitational wave form in the frequency and time domain for an arbitrary trajectory up to 1.5PN order.  They summarize
and generalize the results  for given values of $\ell$, obtained using MPM methods in \cite{Mishra:2015bqa},  to arbitrary orbital number $\ell$. We showed also that the infinite series of tail and tail-of-tail contributions to both, the mass ($U$) and current ($V$) components, of the wave form can be resummed and written in the compact form (\ref{atail}). Interestingly we find, as a new result, that the expressions for the mass and the current tail-of-tail are identical for all multipoles, see the discussion at the end of Section~\ref{sec:wf}.
\\
Finally, we derive a compact and universal formula generalizing the Weinberg result for the scattering amplitude describing the emission of a soft graviton in a two to two collision. Our results resum an infinite series of log-enhanced contributions proportional to $\omega^{n-1} \log^n \omega$ in the soft limit $\omega \to 0$. We find perfect agreement with the probe limit of the conjecture in \cite{Alessio:2024onn} extending the results of \cite{Sahoo:2018lxl,Sahoo:2020ryf}.
\\
The techniques discussed in this paper can be extended in various directions. For instance it should be possible to generalise~\eqref{eq:larger} resumming the instanton contribution of one gauge group also at subleading orders in the coupling of the other gauge group. This would open the avenue to derive, within this framework, scattering waveforms for arbitrary velocities at subleading PM orders. Results in this direction could also be used to refine the analysis of the soft limit presented in Section~\ref{sec:soft}. As we discussed, it is possible to use the approach based on the ${\mathcal N}=2$ quiver gauge theory to reconstruct also the near zone gravitational field: it would be interesting to use the results obtained in this way to study absorption phenomena. Lastly, while in this paper we focused out of simplicity on the Schwarzschild geometry, the technology we discussed can be readily applied also to the Kerr geometry, as already done in~\cite{Fucito:2023afe}. One could study in that case the aspects discussed in this paper as well, including the general form of the tail-of-tail contributions and the log-enhanced soft terms.

\section*{Acknowledgements}
It is a pleasure to acknowledge fruitful scientific exchange with F. Alessio, G. Almeida, M. Bianchi, D. Bini, L. Blanchet, G.~Bonelli, G. Di Russo, P. Di Vecchia, G. Faye, S. Foffa, C. Heissenberg, P. Pani, R. Sturani, A.~Tanzini and G. Veneziano.
F.~Fucito and J.~F.~Morales  thank the MIUR PRIN contract 2020KR4KN2 ``String Theory as a bridge between Gauge Theories and Quantum Gravity'' and the INFN project ST\&FI ``String Theory and Fundamental Interactions'' for partial support.
R.~Russo. is partially supported by the UK EPSRC grant ``CFT and Gravity: Heavy States and Black Holes'' EP/W019663/1 and the STFC grants ``Amplitudes, Strings and Duality'', grant numbers ST/T000686/1 and ST/X00063X/1. No new data were generated or analysed during this study.

\begin{appendix}

\section{The source term  }
\label{apTeu}

In this appendix we review the computation of the source term in (\ref{teukbis}).
 We consider the perturbations of the metric  of spin $s=-2$ defined as\footnote{The different sign with respect to the definition of $\psi_4$ in
\cite{Teukolsky:1972my} arises from the fact that we work in the mostly plus signature. }
  \be
 \psi=r^{4} \psi_4 =r^{4} C_{\mu\nu\sigma \rho} n^\mu\overline{m}^\nu n^\sigma \overline{m}^\rho \label{psi4}\;,
 \ee
 with $C_{\mu\nu\sigma \rho} $ the Weyl tensor perturbation,
 \begin{equation}
n^\alpha = \ft{1}{2 } \left(1,-f(r),0,0\right)\;,\quad  \overline{m}^\alpha=\ft{1}{\sqrt{2}  r }
\left(0,0,1,-{{\rm i} \over \sin\theta} \right)
\end{equation}
and $x^\mu=(t,r,\theta,\phi)$. Note that $\overline{m}^\alpha$ is equal to $e=(0,{\bf e})$ with ${\bf e}$ the polarization vector in (\ref{polarizationvector}) written in Cartesian coordinates.
 The Einstein equations for $\psi$ can be separated at linear order into a radial and an angular part via the ansatz (\ref{psibis}) leading to
 (\ref{teukbis}) with stress energy source
\be
 T_{\ell m}(r) =\pi  \int d\theta d\phi  dt   \sin\theta  S_{\ell m} (\theta) e^{-im\phi+i\omega t}   {\cal T}(x ) ~,
\label{eqsource}
\end{equation}
where $S_{\ell m}(\theta)=Y_{-2}^{\ell m}(\theta, 0)$ , ${\cal T}(x )$ given by  \cite{Mino:1997bx}
\begin{align} \label{eqB}
 {\cal T} (x) & =  2 r^4  L_{-1} L_0
{\cal T}_{nn}   {+} \sqrt{2}\,  r^3 f^2
 J_+\left(  {r^2 \over f}
L_{-1} {\cal T}_{{\bar m}n} \right) \\ &
 {+} \sqrt{2} r \,f^2 \, J_+ \left(
{r^4\over f}    L_{-1}
{\cal T}_{{\bar m}n} \right) {+}  f^2 r J_+ \left[ r^{4}J_+
( r {\cal T}_{{\bar m}{\bar m}}) \right]\nn
  \end{align}
 and
\begin{align}
 {\cal T}_{n n} &= {\cal T}_{\mu \nu} n^\mu n^\nu, \quad {\cal T}_{n \overline{m} } ={\cal T}_{\mu \nu} n^\mu \overline{m}^\nu , \quad
 {\cal T}_{\overline{m } \overline{m}} ={\cal T}_{\mu \nu} \overline{m}^\mu \overline{m}^\nu,   \nn\\
L_s&=\partial_{\theta}{+}{m \over \sin\theta}
 {+}s\cot\theta , ~ L^\dagger_s=\partial_{\theta}{-}{m \over \sin\theta}
 {+}s\cot\theta ,~\nn\\
J_\pm  &= \partial_r{\pm} {i \omega  \over  f  }\;. \label{eqT}
\end{align}
We write
 \bea
 { {\cal T}({\bf x} ) \over r^4 f^2} &=&    L_{-1} L_0\left(
 2 f^{-2} {\cal T}_{nn} \right)  {+}  2 \sqrt{2}\, L_{-1} \left( 2+ J_+ r\right)  f^{-1} {\cal T}_{{\bar m}n}    \nn\\
 &&{+}  \left( J_+^2 r^2 + 2 J_+ r \right)   {\cal T}_{{\bar m}{\bar m}} \label{t22}
 \eea
and use the integration by part identities
\be
\int d\theta  \sin\theta\, S_{\ell m}(\theta)  L_{-s} f(\theta) =\int d\theta  \sin\theta\,  f(\theta) L_{s+1}^\dagger S_{\ell m}(\theta)
\ee
to bring all $\theta$-derivatives of the source to derivatives acting on $S_{\ell m}$.
Plugging (\ref{t22}) into (\ref{rt0bis}) and integrating by parts to bring $r$-derivatives to act on  $\mathfrak{R}_{{\rm in},\ell}$, one finds

\begin{widetext}
\begin{align}
  & Z_{\ell m} =     \int_{r_+}^\infty  \mathfrak{R}_{{\rm in},\ell} (r)     {T_{\ell m }(r) \over  r^4 f^{2}  } dr =  \pi   \int dr d\theta d\phi  dt    e^{-im\phi+i\omega t} \sin\theta  \mathfrak{R}_{{\rm in},\ell} (r)  \left[
  \left(
 2 f^{-2} {\cal T}_{nn} \right)L^\dagger_1  L^\dagger_{2}    \right. \nn\\
 & \quad\quad \left. {+}  2 \sqrt{2}\,   \left( 2+ J_+ r\right)  f^{-1} {\cal T}_{{\bar m}n}   L^\dagger_{2}  {+}  \left( J_+^2 r^2 + 2 J_+ r \right)   {\cal T}_{{\bar m}{\bar m}}
   \right] S_{\ell m} (\theta)
   \\
  &= \pi   \int dr d\theta d\phi  dt    e^{-im\phi+i\omega t} \sin\theta  \left[
  \left(
 2 f^{-2} {\cal T}_{nn} \right)L^\dagger_1  L^\dagger_{2}    {+}  2 \sqrt{2}\,     f^{-1} {\cal T}_{{\bar m}n}  \left( 2- r J_- \right)  L^\dagger_{2}   {+} {\cal T}_{{\bar m}{\bar m}}\left( r^2 J_-^2  - 2 r J_- \right)
   \right]  \mathfrak{R}_{{\rm in},\ell} (r)  S_{\ell m} (\theta)
   \nn\\
& =   \pi   \int  {d\tau \over  r^2   } e^{-im\phi(t)+i\omega t} \Big[
   {2\over  f^2 }  (u \cdot n)^2  L_1^\dagger L_2^\dagger  -{2\sqrt{2} \over  f} (u\cdot \bar m) (u\cdot n)  (2-r J_-)L_2^\dagger + (u\cdot {\bar m})^2 \, \left( r^2 J_-^2  - 2 r J_- \right)  \Big]  S_{\ell m} (\theta)  \mathfrak{R}_{{\rm in},\ell}(r)\;,\nn
\end{align}
with
  \be
u \cdot n =- \ft12 (E+ \partial_\tau r )  \qquad , \qquad  u \cdot \bar{m} = -{{\rm i} J  \over \sqrt{2} r}
\ee
and
\be
   L_1^\dagger L_2^\dagger  S_{\ell m}(\ft{\pi}{2}) =  \sqrt{(\ell-1)\ell(\ell+1)(\ell+2)} \,   {}_0Y_{\ell m}(\ft{\pi}{2})\;,\quad
   L_2^\dagger S_{\ell m}(\ft{\pi}{2}) = -   \sqrt{(\ell-1)(\ell+2) }   \, {}_{-1} Y_{\ell m} (\ft{\pi}{2})\;.
   \ee
\end{widetext}

 \section{Circular orbits}

 Circular orbits are obtained by taking $r(\tau)=r_0$ constant.
The energy, angular momentum and frequency become\footnote{Here we use the expressions written in Schwarzschild coordinates; the waveform~\eqref{eqUc},~\eqref{eqVc} is gauge independent and does not depend on this choice when written in terms of physical quantities.}
 \be
   E= {\mu(1-2 v^2)\over \sqrt{1-3 v^2} } , \quad J={\mu  M  \over v  \sqrt{1-3 v^2} }, \quad  \omega  ={m v^3\over M}
 \ee
 with $v^2 =M/r_0$. We can then use $v$ as the PN expansion parameter. One finds
 \begin{equation}\label{eqUc}\begin{aligned}
 {W}_{\ell m}^U & =  {\mu \, c^0_{\ell m} \, ({\rm i} m v )^\ell} \\ & \Big( 1-v^2 \frac{(4 \ell^3+8 \ell^2+\ell m^2+\ell+9 m^2-3)}{2 (\ell+1) (2
   \ell+3)}  \\
   &   + 2{\rm i} m v^3 \Big[ -\frac{\left( \ell^3+7 \ell^2+12 \ell+8\right)}{ 2\ell (\ell+1) (\ell+2)} \\& - \psi (\ell-1)   + \log(-4 m v^3 )\Big]+\ldots \Big)
 \end{aligned}\end{equation}
and
 \begin{align}\label{eqVc}
&{W}_{\ell m}^V =-{\rm i}  \mu \ell ({\rm i} m v^2)^\ell \, c^1_{\ell m} \Bigg(  1 \\
&  +v^2 \frac{ \left(-4
   l^4-12 l^3-l^2 \left(m^2-3\right)+l \left(34-4
   m^2\right)+24\right)}{2 l \left(2 l^2+7 l+6\right)} \nn \\
& + 2 {\rm i} m v^3 \left[ - \frac{ \left( \ell^3+4 \ell^2-3\ell+2\right) }{2 \ell \left(\ell^2-1\right)} \right. \nn \\ &\quad - \psi(\ell-1)+ \log (-4 mv^3 )\Big]\Bigg)\;.\nn
 \end{align}
The first few terms in the PN expansion are
 \begin{align}
  {W}^{\rm U}_{22}   &=  -2\sqrt{ \ft{\pi}{5}}\,  \mu  v^{2}\,   \left( 1 -\frac{107 v^2}{42} +2 \pi  v^3 +\ldots \right)\;, \nn \\
   {W}^{\rm V}_{21} &=   \frac{ 2 {\rm i}}{3} \sqrt{ \ft{\pi}{5}}\,   \mu v^{2}\,   \left( 1-\frac{17 v^2}{28}+ v^3 \left[\pi -\frac{\rm i}{2}  -2 i \log 2 \right]\right)\;, \nn \\
 {W}^{\rm U}_{33}   &=   -\frac{9{\rm i}}{2} \sqrt{ \ft{\pi}{42}}\,  \mu     v^{3}\,   \Big(  1-4 v^2+ v^3 \Big[3 \pi  -\frac{21  } {5} {\rm i}\nn\\
 &+6 {\rm i}  \log  \left(\frac{3}{2}\right)\Big] +\ldots \Big)\;, \nn\\
  {W}^{\rm V}_{32}   &=  - \frac{2{\rm i}}{3} \sqrt{ \ft{\pi}{7}}\,  \mu  v^{3}\,   \left(  1-\frac{193 v^2}{90} +   (2 \pi -3  {\rm i}) v^3+\ldots \right)\;, \nn \\
 {W}^{\rm U}_{31}   &=     \frac{{\rm i}}{6} \sqrt{ \ft{\pi}{70}}\,    \mu   v^{3}\,   \Big(  1-\frac{8 v^2}{3}\nn\\
   &+ v^3 \Big[-\frac{7 i}{5}+\pi -2 i \log
   (2)\Big] +\ldots  \Big)\;,
\end{align}
 in agreement\footnote{As done in \cite{Blanchet:2013haa}, we redefined the retarded time so as to eliminate all constants in the $\ell=2$ tail~\eqref{aulm}, except for the term of ${\rm i}\pi/2$ which cannot be reabsorbed as it is imaginary. This yields an $\ell$-independent shift for the other modes.} with (488)-(495) in \cite{Blanchet:2013haa}.

\section[N=2 supersymmetric Yang-Mills gauge theory and gravity]{$N=2$ supersymmetric Yang-Mills gauge theory and gravity}
\label{appendixC}

In this appendix we briefly review the connection between the solutions of (\ref{teukbis}) and the five-point degenerate correlator in Liouville theory which, in turn via the AGT correspondence, is tantamount to the partition function of a $N=2$ supersymmetric Yang-Mills quiver gauge theory with gauge group $SU(2)^2$. We refer the reader to \cite{Consoli:2022eey,Fucito:2023afe} for a more detailed presentation.

 To build the dictionary between these two theories, we bring the homogeneous part of (\ref{teukbis}) to the Schr\"odinger like form
 \be
\left[ {d^2\over dz^2} +Q(z) \right] \Psi(z) = 0\;,  \label{can}
\ee
with
\be
z = {2M\over r}\;,  \quad    R(r) ={(1-z)^{1\over 2} \over z^2  }   \Psi(z) \label{rzdic}
\ee
and
\be
Q=\frac{4 (\ell^2\!+\!\ell) z^2(z\!-\!1) \!+\! 16 M^2 \omega ^2\!- \! 3 z^4\!+\!\ii M z \omega(48z\!-\!32)}{4 (z-1)^2 z^4}. \label{qr}
\ee
The same equation arises as a conformal Ward identity satisfied by a degenerate five point correlator in Liouville theory.
By comparing~\eqref{qr} with the parametrization of $Q$ given in Eq.~(3.2) of~\cite{Fucito:2023afe} in terms of $p_0$, $k_0$, $c$, $u$ and $x$, one obtains the following dictionary with the gravity variables 
\bea
x  &=& 4 i   \omega M  \;, \label{dicrsch}\\
 u &=&  \mathfrak{a}^2 -  {x}  \partial_{x} {\cal F}_{\rm inst} (\mathfrak{a},x)  = (\ell{+}\ft12)^2 {-}2 i M \omega {-}20 \omega^2 M^2\;,   \nn \\
c &=&  {-}2{-}2 i M \omega \;,  ~~  p_0 =  -1   \;, ~~  k_0 =
{-}1{+} 2 i M \omega\;.  \nn
\eea
Expanding ${\cal F}_{\rm inst}$ in $x$ and inverting (\ref{dicrsch}) one finds
\be
\mathfrak{a}=\widehat\ell+\ft12
\ee
where $\widehat\ell$ is given by (\ref{lhatl}).
 The two independent solutions $\Psi_\alpha(z)$ can be written as
\begin{align}
\Psi_\alpha(z) & =  \lim_{b \to 0}\, e^{{x\over 2z} }  \,  z^{{1\over 2} +\alpha \mathfrak{a}}     \,
\left(1{-} z  \right)^{ (2k_{0}{-}1 {-} b^2)( 2 k_{\rm deg} {+} 1 {+}  b^2 ) \over 2 b^2  }  \nn\\
& \quad {Z_{\rm inst}{}_{p_0}{}^{k_0}  {}_{\mathfrak{a}^{-\alpha}}{}^{ k_{\rm deg}}  {}_{\mathfrak{a}}  {}_c  (z,\ft{x}{z} ) \over
	Z_{\rm inst}{}_{p_0}{}^{k_0} {}_{\mathfrak{a}}   {}_c (x)   }\;,  \label{psidef}
\end{align}
are written in terms of the instanton partition functions of the corresponding $SU(2)$ and $SU(2)^2$ gauge theories, which are respectively
\begin{widetext}
\begin{align}
Z_{\rm inst}{}_{p_0}{}^{k_0} {}_{ \mathfrak{a} c} (x) & = e^{ {\cal F}_{\rm inst} (\mathfrak{a},x) \over b^2} = \sum_{ W}  x^{ |W |}
	 \frac{ z^{\rm bifund}_{\emptyset,W} (p_0,\mathfrak{a},-k_0)    z^{\rm hyp}_{W} (\mathfrak{a},-c)
	 }{     z^{\rm bifund}_{W,W} (\mathfrak{a},\mathfrak{a},\ft{b^2+1}{2}) } \,  \label{zinstG2} \\
	Z_{\rm inst}{}_{p_0}{}^{k_0}  {}_{\mathfrak{a}^{-\alpha} } {}^{k_1} {}_{ \mathfrak{a} c} (q_1,q_2) & =
	 \sum_{ Y,W}  q_1^{  |Y| } q_2^{ |W |}
	 \frac{ z^{\rm bifund}_{\emptyset,Y} (p_0,\mathfrak{a}^{-\alpha} ,-k_0) z^{\rm bifund}_{Y,W} (p_1,\mathfrak{a},-k_1)   z^{\rm hyp}_{W} (\mathfrak{a},-c)
	 }{  z^{\rm bifund}_{Y,Y} (\mathfrak{a}^{-\alpha},\mathfrak{a}^{-\alpha},\ft{b^2+1}{2})   z^{\rm bifund}_{W,W} (\mathfrak{a},\mathfrak{a},\ft{b^2+1}{2}) }\;,
  \label{inst1}\end{align}
  \end{widetext}
with $q_1=2  M/r$ and $q_2=2 i\omega r$. The sums in (\ref{inst1}) run over the pairs of Young tableau $\{Y_\pm \}$, $\{W_\pm \}$ while $|Y|$, $|W|$ denote the total number boxes in each pair.
The functions  $z^{\rm bifund}_{Y,W }$, $z^{\rm hyp}_{Y_{2} }$ represent the contributions of a hypermultiplet transforming in the bifundamental and fundamental representation of the gauge group and are given by a product over the boxes of the Young tableau
\begin{align}
	& z^{\rm bifund}_{\Lambda, \Lambda' }( p , p' , m ) =  \prod_{\beta,\beta'}
	\left\{\! \prod_{(i,j)\in \Lambda_\beta} \!\!\!\! \left[  E_{\Lambda_\beta,\Lambda'_{\beta'}}\!(\beta p{-}\beta' p',i,j)  {-} m  \right] \right. \nn \\
& \left.\prod_{(i',j')\in \Lambda'_{\beta'}}\!\!\! \left[  { -}E_{\Lambda'_{\beta'},\Lambda_\beta}(\beta' p' {-}\beta p,i',j')  {-} m  \right]\right\}\;,  \nn\\
&z^{\rm hyp}_{\Lambda }(   \mathfrak{a} ,m ) =   \prod_{\beta}
	    \prod_{(i,j)\in \Lambda_\beta}  \left[ { -}E_{\Lambda_\beta,\emptyset}(\beta \mathfrak{a},i,j) { +}m  \right]\;,
\label{inst2}
\end{align}
with $\beta,\;\beta'=\pm 1$ and
\begin{equation}
E_{\Lambda, \Lambda'}(x,i,j) = x- (\lambda_{ \Lambda' j}^T-i) + b^2 (\lambda_{ \Lambda i}-j+1) -\ft{b^2+1}{2}\;,
\label{inst3}
\end{equation}
where $\lambda_{ \Lambda i}$ is the number of boxes in the $i$-th row and $\lambda_{ \Lambda j}^T$ is the number of boxes in the $j$-th column of the tableau $\Lambda$.
The first instanton contributions to the partition function of the $SU(2)$ gauge theory is
\begin{align}
{\cal F}_{\rm inst} & =  \lim_{b\to 0} b^2   \ln \Big[  Z_{\rm inst}{}_{p_0}{}^{k_0} {}_{\mathfrak{a}}   {}_c (x) \Big]
\nn\\
& =  \frac{x}{2} \Big[ (1+c-2k_0) -{4c \left( k_0^2- p_0^2\right) \over 1-4 \mathfrak{a}^2} \Big] +  \ldots\;,   \label{u4}
\end{align}
 while the quiver partition function is given by the double instanton series
\begin{align}
&\Psi_\alpha(z) =  z^{{1\over 2} +\alpha \mathfrak{a}} \left[  1+z \frac{ k_0^2- p_0^2+
	\mathfrak{a}^2-\ft14}{1+2 \alpha  \mathfrak{a}} -\frac{c {x} }{z \left(1-2 \alpha  \mathfrak{a}\right)} \right . \nn\\
&   \left.  -
c {x} \frac{4 \!\left(k_0^2-p_0^2 \right) \left(2 \alpha  \mathfrak{a}\! + \! 3\right)-(3\! - \! 2\alpha \mathfrak{a})(1 \! + \! 2\alpha \mathfrak{a})^2  }{4 \left(1-2 \alpha  \mathfrak{a}\right) \left(2
	\alpha  \mathfrak{a}+1\right){}^2}  \right] +\ldots\;. \label{psialpha}
\end{align}
The general solution of the homogenous Teukolsky equation is given by a linear combination of the two independent solutions $  z^{-2} (1-z)^{1\over 2}  \Psi_\alpha(z) $.  In particular, a solution
$R_{\rm in}(z)$ satisfying at the horizon  incoming boundary  conditions
\be
R_{\rm in}(z) \underset{z\to 1}{\approx} (1-z)^{1-k_0}
\ee
is proportional to the linear combination
\be
R_{\rm in} (z) =  z^{-2} (1-z)^{1\over 2}  \sum_{\alpha = \pm} F^{-1}_{- \alpha }  \Psi_\alpha(z)\,.
\ee
 The overall normalization is determined by imposing the asymptotic behaviour (\ref{rinf}). The asymptotic form of  $R_{\rm in}(z)$ can be obtained with the help of the CFT
 braiding and fusion relations. One finds
 \be
 R_{\rm in} (z)
\underset{z\to 0}{\approx}   \sum_{\alpha'} B^{\rm in}_{\alpha'} \, e^{{\alpha' {x} \over 2 z } } z^{ -1+\alpha' c}  \label{connection}
\ee
with
\be
B^{\rm in}_{\alpha'}= \sum_{\alpha}    F^{-1}_{- \alpha}
				  B^{\rm conf}_{\alpha  \alpha'}  \,   x^{ -{1\over 2} -\alpha' c+\alpha \mathfrak{a}}e^{-\frac{1}{2}  (\alpha' \partial_c +\alpha \partial_{\mathfrak{a}}){\cal F}_{\rm inst}(x)}\,.
				    \label{calphap}
\ee
 Imposing (\ref{rinf}) one finds
    \be
\mathfrak{R}_{{\rm in},\ell}(r) ={ R_{\rm in} (z)\over 4 {\rm i} \omega M B^{\rm in}_-} =\mathfrak{R}_\mathfrak{a}(r) +\mathfrak{R}_{-\mathfrak{a}}(r)
\ee
with
\be
 \mathfrak{R}_{\mathfrak{a}}(r)  =  {z^{-2} (1-z)^{1\over 2} F^{-1}_{- - }  \Psi_-(z)     \over   \sum_{\alpha} F^{-1}_{- \alpha}
B^{\rm conf}_{\alpha -}   x^{{1\over 2}{+}c{+}\alpha \mathfrak{a}}  e^{{ 1\over 2}  ( \partial_c-\alpha \partial_{\mathfrak{a}}) {\cal F}_{\rm inst}    }  }
\ee
and
\begin{widetext}
\begin{equation}
F^{-1}_{\alpha'' \alpha }  = \frac{\Gamma (1+2  \alpha'' k_0)
	\Gamma \left(-2  \alpha \mathfrak{a}\right)}{\Gamma \left(\frac{1}{2}+\alpha'' k_0+p_0-\alpha \mathfrak{a}
	\right) \Gamma \left(\frac{1}{2}+\alpha'' k_0-p_0-\alpha \mathfrak{a}\right)} \;, \quad		
B^{\rm conf}_{\alpha \alpha'}  =  e^{\frac{{\rm i} \pi (1{-}\alpha' )}{2}   (\frac{1}{2} -c{-}\alpha a )}   \frac{   \Gamma \left(1-2  \alpha \mathfrak{a}\right) }
{\Gamma \left(\frac{1}{2} -\alpha \mathfrak{a}-\alpha' c \right) }\;.  \label{braidingBconf}
\end{equation}
Finally $\mathfrak{a}$ can be obtained by inverting (\ref{u4}) and using (\ref{dicrsch}) leading to
\begin{equation}
\mathfrak{a}=\ell +\ft{1}{2}-2\omega^2 M^2\frac{\left(15 \ell^4+30 \ell^3+28 \ell^2+13 \ell+24\right) }{\ell (\ell+1) (2 \ell-1) (2 \ell+1) (2 \ell+3)}+ O\left((\omega M)^4\right)\;. \label{p2finalapp}
\end{equation}
\end{widetext}

\end{appendix}

\providecommand{\href}[2]{#2}\begingroup\raggedright\endgroup

 \end{document}